\documentclass[10pt]{iopart}

\usepackage{iopams}
\usepackage{amssymb}
\usepackage{amsfonts}
\usepackage{amstext}
\usepackage{graphicx}
\usepackage{setstack}
\usepackage{amscd}
\usepackage{mathrsfs}
\usepackage{stmaryrd}
\usepackage{bbm}
\usepackage{color}

\newcommand{\IM}{\Im\mathrm{m}}
\newcommand{\llangle}{\langle \hspace{-0.2em} \langle}
\newcommand{\rrangle}{\rangle \hspace{-0.2em} \rangle}

\newcommand{\rrrangle}{\rangle \hspace{-0.2em} \rangle \hspace{-0.2em} \rangle}
\newcommand{\id}{\mathrm{id}}

\newcommand{\Sp}{\mathrm{Sp}}
\newcommand{\mod}{\ \mathrm{mod}\ }

\newcommand{\Env}{\mathrm{Env}}

\begin{document}

\title[D-brane dynamics with fluctuations]{Chaos, decoherence and emergent extradimensions in D-brane dynamics with fluctuations}

\author{David Viennot \& Lucile Aubourg}
\address{Institut UTINAM (CNRS UMR 6213, Universit\'e de Bourgogne-Franche-Comt\'e, Observatoire de Besan\c con), 41bis Avenue de l'Observatoire, BP1615, 25010 Besan\c con cedex, France.}

\begin{abstract}
  We study, by using tools of the dynamical system theory, a fermionic string streched from a non-commutative D2-brane (stack of D0-branes in the BFSS model) to a probe D0-brane as a quantum system driven by a chaotic system, the (classical and quantum) chaos being induced in the D2-brane dynamics by quantum fluctuations. We show that this dynamics with fluctuations induces a decoherence phenomenon on the reduced density matrix of the fermionic string which is characteristic of the chaotic behaviour since it presents an horizon of coherence. Moreover we show that this dynamics is associated with an invariant torus which involves extradimensions emerging from the fluctuations for the viewpoint of the fermionic string, extending a three dimensional space by six compact dimensions. The situation studied can be considered as a model of qubit (supported by the fermionic string) in interaction with a quantum black hole (modelized by the non-commutative D2-brane).
\end{abstract}


\section{Introduction}
In a recent paper \cite{Viennot}, we have studied the decoherence phenomenon induced on a qubit (supported by the spin of a fermion) falling into a black hole. This study has needed some strong semi-classical approximations in quantum field theory on curved space-time (only first quantization of the fermion, localization of the qubit (WKB approximation), adiabatic approximation of the dynamics). In the present paper, we want to consider this problem in a ``quantum gravity context''. An interesting model of quantum black holes arises in the matrix model with D-branes \cite{Asplund, Berenstein, Aoki}. In the BFSS (Banks-Fischler-Shenker-Susskind) matrix theory \cite{BFSS}, a stack of D0-branes forms a noncommutative D2-brane represented by Hermitian matrices (the noncommutative coordinates on the D2-brane).  The initial conditions on these matrices fix the topology of the D2-brane (as for example a fuzzy sphere) and adding small initial fluctuations induces thermalization of the system in a state corresponding to a black hole horizon. By adding a fermionic string stretched from the D2-brane to a probe D0-brane, as for example in \cite{Berenstein}, we have a qubit model supported by the spin inner degree of freedom of the fermionic string. It has been proved \cite{Asano, Gur, Hanada} that under these conditions the matrix brane dynamics involves chaotic motions. The fermionic string is then driven by a chaotic system. In previous papers \cite{Viennot2, Aubourg, Aubourg2} we have shown that quantum systems driven by chaotic flows are distinguished from quantum systems driven by random noises by a phenomenon called horizon of coherence, i.e. a duration at the beginning of the dynamics in which the quantum coherence is sustained before to fall with an increase of the entropy. In the present paper, we want to study this phenomenon for a fermionic string in contact with a D2-brane. Moreover we have proposed in \cite{Viennot3} a natural mathematical approach to study quantum systems driven by chaotic flows, based on the Schr\"odinger-Koopman formalism \cite{Sapin, Jauslin, Gay}. This formalism is a generalization of the Floquet theory for which we have shown \cite{Viennot4} that it has similarities with the classical string theory. In particular, a closed extradimension must be added to the control manifold (playing the role of the space) in the context of this Floquet theory. In this paper, we want to show that by a similar mechanism, compact extradimensions emerge from the quantum chaotic fluctuations from the viewpoint of the fermionic string treated by the Schr\"odinger-Koopman approach. For this reason, at the starting point we consider a matrix model in a three dimensional space, the other compact six dimensions of the BFSS model emerging from the fluctuations in our approach. Note that the reduction to three dimensions, can also be interpreted as an orbifold reduction as in \cite{Berenstein}.\\
This paper is organized as follows. Section 2 presents elementary ingredients of the BFSS matrix theory. The different models used for the numerical simulations are also presented. Section 3 studies the dynamics of the quantum fluctuations, in particular we show the existence of an invariant torus in the phase space of the matrices modelizing the noncommutative D2-brane. Section 4 presents the main results of this paper concerning the dynamics of the fermionic string driven by the D2-brane: a quantum chaos signature in its spectrum, an horizon of coherence in the evolution of its reduced density matrix and the emergence of the compact extradimensions from the fluctuations. Throughout this paper, our results are enlighten by numerical simulations. In particular, we drawn the geometry of the compact extradimension manifold emerging in our approach.\\

\textit{Throughout this paper, we consider the unit system such that $\hbar=c=G=1$ ($\ell_P=t_P=m_P=1$ for the Planck units)}.

\section{The BFSS model}
In this section we present the elementary ingredients of the BFSS theory, to more details the reader can see \cite{BFSS, Sochichiu, Zarembo}.\\

We consider a stack of $N$ D0-branes in a 3D-space, represented by Hermitian matrices $X_i \in \mathfrak M_{N \times N}(\mathbb C)$ for $i=1,2,3$ (noncommutative coordinates for the D2-brane formed by the stack). Intuitively, $X^i = \left(\begin{array}{cc} x^i_1 & s^i_{12} \\ \overline s^i_{12} & x^i_2 \end{array} \right)$ represents a stack of two D0-branes with coordinates in the target space $\mathbb R^3$ $\{x^i_1\}_i$ and $\{x^i_2\}_i$, linked by a bosonic string of oscillation radii $\{|s^i_{12}|\}_i$ ($\arg s^i_{12}$ is the phase for the oscillation in the $i$-direction).\\
We consider also a probe D0-brane of coordinates $x_i \in \mathbb R$ and a fermionic string linking the D2-brane to the probe brane described by a state:\\ $|\psi \rrangle = |\uparrow \rangle \otimes \left(\begin{array}{c} \psi^\uparrow_1 \\ \vdots \\ \psi^\uparrow_N \end{array} \right)+ |\downarrow \rangle \otimes \left(\begin{array}{c} \psi^\downarrow_1 \\ \vdots \\ \psi^\downarrow_N \end{array} \right) \in \mathbb C^2 \otimes \mathbb C^N$. We can interpret $\psi^\alpha_a$ as $\psi^\alpha(x_a)$ ($x_a$ being the pseudo-position of the $a$-th D0-brane of the stack): the spatial delocalisation of quantum particles is replaced by the quantum superposition of $N$ attachment points for the string. The inner degree of freedom modelized by the Hilbert space $\mathbb C^2$ (with canonical basis $(|\uparrow \rangle, |\downarrow \rangle)$ is the spin of the fermionic string.\\
The total system is represented by the matrices:
\begin{equation}
  \mathbf X_i = \left(\begin{array}{cc} X_i & |\delta x_i\rangle \\ \langle \delta x_i| & x_i \end{array} \right) ; \qquad \Psi = \left(\begin{array}{cc} 0 & |\psi\rrangle \\ 0 & 0 \end{array} \right)
\end{equation}
where $|\delta x_i \rangle \in \mathbb C^N$ represents quantum fluctuations of the vacuum creating and annihilating bosonic strings linking the stack (the D2-brane) to the probe D0-brane. $|\delta x_i(t=0)\rangle$ is randomly choosen following a gaussian law.\\

The dynamics of the system is governed by the Lagrangian densities:
\begin{equation}
  \mathscr L_{boson} = \frac{1}{2} \dot {\mathbf X}^i \dot {\mathbf X}_i + \frac{1}{4} [\mathbf X_j,\mathbf X^i][\mathbf X^j,\mathbf X_i]
\end{equation}
\begin{equation}
  \mathscr L_{fermion} = -\imath \Psi^\dagger \dot \Psi + \Psi^\dagger \sigma^i \otimes [\mathbf X_i,\Psi]
\end{equation}
involving the following equation of motion:
\begin{equation}
  \label{equX}
  \ddot {\mathbf X}_i - [\mathbf X_j,[\mathbf X_i,\mathbf X^j]] = 0
\end{equation}
with the Gauss constraint $[\dot {\mathbf X}_i,\mathbf X^i] = 0$,
\begin{equation}
  \label{equSchro}
 \imath |\dot \psi \rrangle = \sigma^i \otimes(X_i-x_i) |\psi \rrangle
\end{equation}
$\{\mathbf X_i\}_i$ obeys then to a matrix classical wave equation ($[\mathbf X_j,[\mathbf X_i,\mathbf X^j]]$ taking the role of a noncommutative Laplacian). The string state $|\psi \rrangle$ obeys to a Schr\"odinger equation with quantum Hamiltonian $H^{eff}(t) = \sigma^i \otimes(X_i(t)-x_i(t))$ \cite{Berenstein, Badyn}.\\

The initial conditions concerning $X_i(t=0)$ fix an initial topology for the noncommutative D2-brane. Usually, the works concerning such a model use a fuzzy sphere at $t=0$. In order to test the independance of our analysis from the initial topology we consider also a fuzzy ellipsoid, a fuzzy cylinder, a fuzzy torus and two noncompact noncommutative manifolds, a noncommutative plane and a fuzzy hyperboloid. The characteristics of these models are presented table \ref{NCmanifold}.
\begin{table}
  \caption{\label{NCmanifold} Noncommutative manifolds used as initial conditions for the D2-brane (see also \cite{Sykora}). For the numerical simulations, the choosen parameters of the models are $\alpha=1.5$, $\delta=0.4$ and $k=12.5$.}
    \begin{tabular}{l||l|l}
      {\it manifold} & {\it algebra} & {\it coordinates} \\
      \hline \hline
      Fuzzy sphere & \begin{minipage}{4cm} $j$-representation of $\mathfrak{su}(2)$ ($N=2j+1$)\\ $[J_i,J_j] = \imath \epsilon_{ijk} J_k$ \end{minipage} & \begin{minipage}{4cm} $X_i = \frac{1}{j} J_i$ \end{minipage} \\
      \cline{1-1} \cline{3-3}
      Fuzzy ellipsoid & & \begin{minipage}{4cm} $X_1=\frac{2}{j} J_1$, $X_2=\frac{2}{j} J_2$, $X_3=\frac{1}{j} J_3$ \end{minipage}  \\
      \hline
      Fuzzy cylinder & \begin{minipage}{4cm} $j$-representation of $\mathfrak{iso}(2)$ ($N=2j+1$)\\ $[C,V]=V$ \end{minipage} & \begin{minipage}{4cm} $X_1 = \frac{1}{j} \frac{V+V^\dagger}{2}$, $X_2 = \frac{1}{j} \frac{V-V^\dagger}{2\imath}$, $X_3 = \frac{1}{j} C$. \end{minipage}\\
      \hline
      Fuzzy torus & \begin{minipage}{4cm} \scriptsize $C = \sum_{n=0}^{N-1} C_n |n\rangle \langle n|$\\ $V=\sum_{n=0}^{N-1} V_n |n\rangle \langle n+1|$\\ with $|N\rangle = |0\rangle$, $C_n = \frac{1}{N\alpha} \sin(\frac{2\pi (n+\delta)}{N})$ and $V_n = \sqrt{\frac{\cos(\frac{2\pi (n+\delta+0.5)}{N})}{N\alpha\cos(\pi/N)}+\frac{1}{N}}$ ($\alpha \in \mathbb R^+$, $\delta \in [0,1]$). \end{minipage} &  \begin{minipage}{4cm} $X_1 = \frac{V+V^\dagger}{2}$, $X_2 = \frac{V-V^\dagger}{2\imath}$, $X_3 = C$. \end{minipage}\\
      \hline
      NC plane & \begin{minipage}{4cm} $\mathfrak{ccr}$ algebra ($N=+\infty$)\\ $[a,a^+]=\id$ \end{minipage} & \begin{minipage}{4cm} $X_1 = \frac{a+a^+}{2}$, $X_2 = \frac{a-a^+}{2\imath}$, $X_3=0$ \end{minipage} \\
      \hline
      Fuzzy hyperboloid & \begin{minipage}{4cm} $k$-representation of $\mathfrak{su}(1,1)$ ($N=+\infty$)\\ $[K_1,K_2]=-\imath K_3$, \\$[K_3,K_1]=\imath K_2$, \\$[K_2,K_3]=\imath K_1$ \end{minipage} & \begin{minipage}{4cm} $X_i = \frac{1}{k} K_i$ \end{minipage}
    \end{tabular}
\end{table}
Numerical simulations for the integration of eq. \ref{equX} are realized by using a leap-frog integrator as in \cite{Asplund}. For the non-compact manifolds, a cutoff in the basis is realized which needs a numerical correction to avoid numerical artefacts (see \ref{AppA}). The integration of the Schr\"odinger equation \ref{equSchro} is realised by a split operator method for the evolution operator: $U(t_n,0) \simeq e^{-\imath H^{eff}(t_n) \Delta t}...e^{-\imath H^{eff}(0) \Delta t}$ for a partition $0<t_1<...<t_n$ of step $\Delta t$ equal to the one used in the leap-frog integrator.

\section{Dynamics of the quantum fluctuations}
\subsection{D-brane equation with fluctuations}
\begin{figure}
  \begin{center}
    \includegraphics[width=6.4cm]{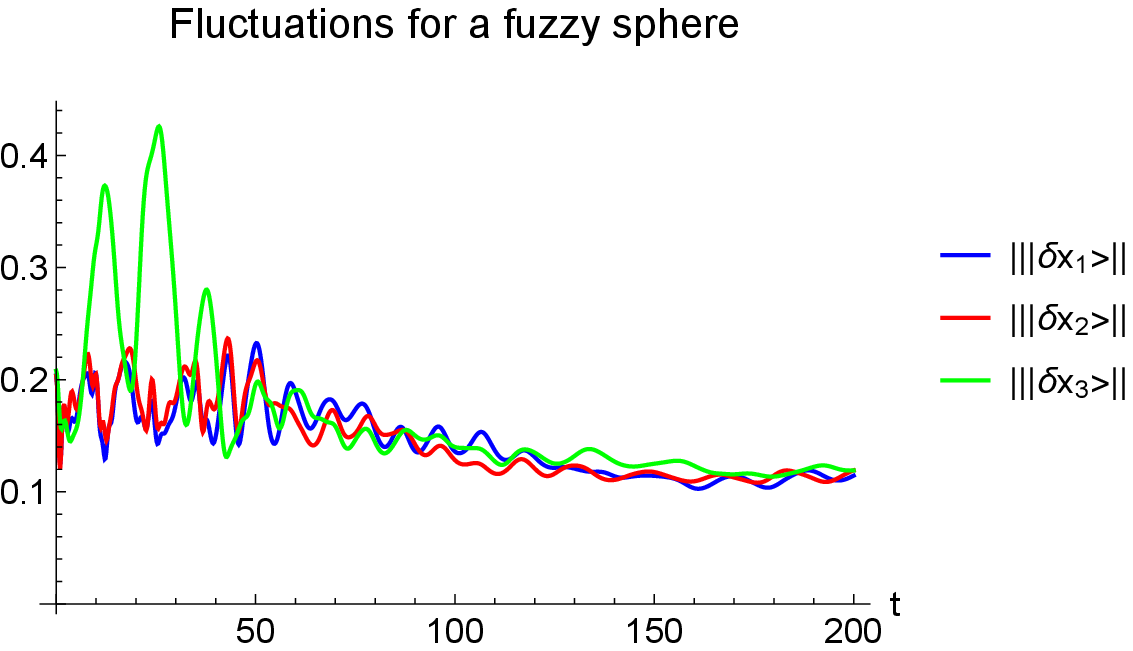} \includegraphics[width=6.4cm]{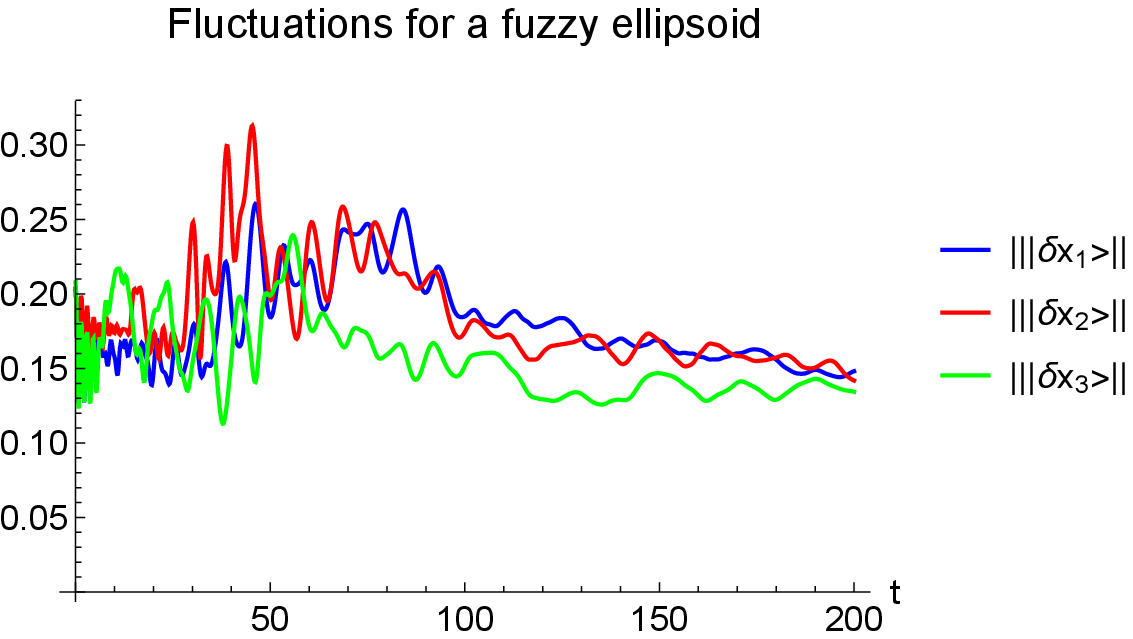}\\
    \includegraphics[width=6.4cm]{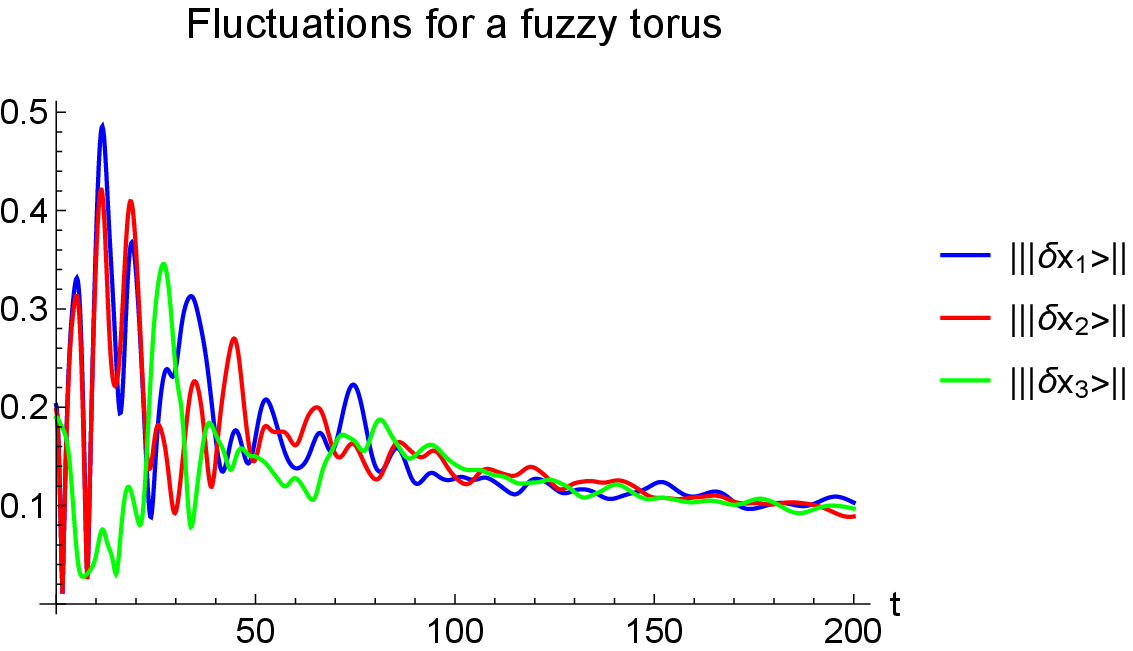} \includegraphics[width=6.4cm]{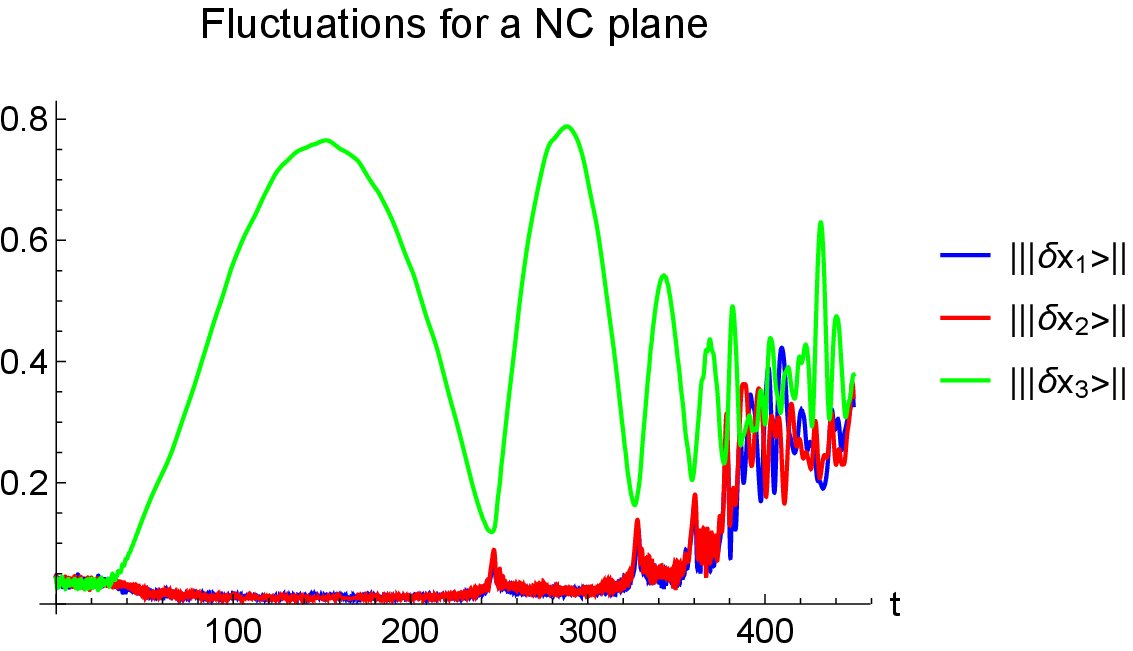}
    \caption{\label{fluctuations} Evolution of the amplitudes of the fluctuations $|\delta x_i(t) \rangle$, for the fuzzy sphere, ellipsoid, torus and the noncommutative plane. Initial conditions for $|\delta x_i(t=0) \rangle$ are choosen following a gaussian law with $\sigma=0.01$. The number of D0-branes is $N=201$ (for the noncommutative plane the cutoff is $N=11$) (simulations show that the thermalization duration grows with $N$).}
  \end{center}
\end{figure}

 The equation $\ddot{\mathbf X}_i - [\mathbf X_j,[\mathbf X_i,\mathbf X^j]]=0$  corresponds to the set of coupled equations:
\begin{eqnarray}
  \ddot X_i & = & [X_j,[X_i,X^j]] \nonumber \\
  & & + [X_j,|\delta x_i\rangle \langle \delta x^j|-|\delta x^j\rangle \langle \delta x_i|] - \{|\delta x_j\rangle \langle \delta x^j|,X_i-x_i\} \nonumber \\
  & & + |\delta x_j\rangle \langle \delta x_i|(X^j-x^j) - (X^j-x^j)|\delta x_i\rangle \langle \delta x_j|
\end{eqnarray}
\begin{equation}
  \ddot x_i = 2\langle \delta x_j|(X_i-x_i)|\delta x^j\rangle -2\Re\mathrm{e}(\langle \delta x_j|(X^j-x^j)|\delta x_i\rangle)
\end{equation}
\begin{eqnarray} \label{equfluct}
  |\delta \ddot x_i \rangle & = & - |X-x|^2|\delta x_i\rangle -([X_i,X^j]-(X^j-x^j)(X_i-x_i))|\delta x_j \rangle \nonumber \\
  & & -\langle \delta x^j|\delta x_j\rangle |\delta x_i\rangle +(2\langle \delta x_i|\delta x^j\rangle - \langle \delta x^j|\delta x_i\rangle)|\delta x_j\rangle
\end{eqnarray}
where $|A|^2 = A_i A^i$ for a set of matrices $\{A_i\}_{i=1,2,3}$ and $\{\cdot,\cdot\}$ denoting the anticommutator. We are interested by the last equation which governs the evolution of the fluctuations. Some examples are drawn figure \ref{fluctuations}.
We see that the fluctuations evolve until to reach a plateau corresponding to the thermalization of the system (the system with the fluctuations has reached a steady state). The existence of this long term stability suggests that the dynamical system described by $(|\delta x_i\rangle)_i$ presents an invariant manifold.    

\subsection{Existence of an invariant torus} \label{invtorus}
We consider the nonlinear equation \ref{equfluct} where we treat $\{X_i\}_i$ and $\{x_i\}_i$ as being constant (the variations of these quantities are supposed very slow compared to the variations of $\{|\delta x_i \rangle\}_i$ and we consider them as being frozen in a first time).\\
We restrict our attention to the case $N=1$ which is instructive for our discussion, $|\delta x_i \rangle$ is then reduced to be a complex scalar $\delta x_i \in \mathbb C$ obeing to the nonlinear equation:
\begin{equation}
  \label{eqsd}
  \delta \ddot x_i = -\|X-x\|^2 \delta x_i + (X^j-x^j)(X_i-x_i)\delta x_j - 2 \overline{\delta x^j} \delta x_j \delta x_i + 2 \overline{\delta x_i} \delta x^j \delta x_j
\end{equation}
We can rewritte this equation as
\begin{equation}
  \left\{ \begin{array}{rcl} \delta \dot x_i & = & F_i(\delta p) = \delta p_i \\ \delta \dot p_i & = & F_{\hat i}(\delta x) \end{array} \right.
\end{equation}
where $\delta p_i \in \mathbb C$ is the momentum associated with $\delta x_i$, $F_{\hat i}(\delta x)$ being the right hand side of the equation \ref{eqsd}. We adopt the notations such that the indices $i$, $\hat i$, $\bar i$ and $\tilde i$ respectively stand for $\delta x_i$, $\delta p_i$, $\overline{\delta x_i}$ and $\overline{\delta p_i}$. These equations define a dynamical system in the phase space $\Gamma = \mathbb C^6 \simeq \mathbb R^{12}$. It is clear that $\delta x_i = 0$ and $\delta p_i = 0$ ($\forall i$) is a fixed point of the dynamical system. Since the quantum fluctuations are small, the starting point of the dynamical system is in the neighbourhood of $0$. The Jacobian matrix of the dynamical system is
\begin{eqnarray}
  {\partial F_{\hat i}}^j  =  \frac{\partial F_{\hat i}}{\partial \delta x_j} & = & -\|X-x\|^2\delta^i_j + (X_i-x_i)(X_j-x_j) \nonumber \\
  & & \quad - 2 \overline{\delta x^j} \delta x_i - 2 \overline{\delta x^k}\delta x_k \delta^i_j + 4 \overline{\delta x_i} \delta x^j \\
  {\partial F_{\hat i}}^{\bar j} = \frac{\partial F_{\hat i}}{\partial \overline{\delta x_j}} & = & -2\delta x_j\delta x_i + 2 \delta x^k \delta x_k \delta^i_j \\
  {\partial F_{i}}^{\hat j} = \frac{\partial F_{i}}{\partial \delta p_j} & = & \delta^i_j
\end{eqnarray}
and ${\partial F_{i}}^{j} = {\partial F_{\hat i}}^{\hat j} = {\partial F_{i}}^{\tilde j} = {\partial F_{i}}^{\bar j} = {\partial F_{\hat i}}^{\tilde j} = 0$. It follows that the Jacobian matrix at $0$ (in the representation $(i,\hat i,\bar i, \tilde i)$) is
\begin{equation}
  \partial F_{|(\delta x,\delta p)=0} = \left(\begin{array}{cc} S & 0 \\ 0 & S \end{array} \right)
\end{equation}
with
\begin{equation}
  S = \left(\begin{array}{cc} 0 & -I \\ \id_3 & 0 \end{array} \right)
\end{equation}
where $I$ is the following ``inertia matrix'':
\begin{equation}
  I = \left(\begin{array}{ccc} Y_2^2+Y_3^2 & -Y_1Y_2 & -Y_1Y_3 \\ -Y_1Y_2 & Y_1^2+Y_3^2 & -Y_2Y_3 \\ -Y_1 Y_3 & -Y_2 Y_3 & Y_1^2+Y_2^2 \end{array} \right)
\end{equation}
with $Y_i = X_i-x_i$. $\Sp(S) \subset \imath \mathbb R$, indeed let $\lambda$ be an eigenvalue of $S$: $Se = \lambda e$. $e = (\mathring e,\hat e)$ with $\mathring e, \hat e \in \mathbb C^3$. The eigenequation is equivalent to $\left\{ \begin{array}{rcl} - I \hat e & = & \lambda \mathring e \\ \mathring e & = & \lambda \hat e \end{array} \right.$, and then $-I \hat e = \lambda^2 \hat e$. $\lambda^2 \in \Sp(-I)$, but $I \geq 0$ and then $\lambda^2 \leq 0 \iff \lambda \in \imath \mathbb R$. $S$ being real, its spectrum contains three couples of conjugated purely imaginary eigenvalues.\\
The 12 local Lyapunov values of the dynamical system in the neighbourhood of $0$ being purely imaginary, the fixed point is simply stable, the evolution of $\delta x$ is then quasi-periodic with 3 fundamental frequencies (the eigenvalues of $\sqrt I$) and the phase trajectory in $\Gamma$ is wrapped around an invariant 6-torus $\mathbb T^6$. By the KAM (Kolmogorov-Arnold-Moser) theorem \cite{Goldstein}, the torus is stable under the perturbations induced by the fluctuations of $X$ and $x$, and it is slowly deformed with the global evolutions of $X(t)$ and of $x(t)$.\\

Now we consider the case $N>1$. By a similar approach, the Jacobian matrix at $0$ of the dynamical system governing the quantum fluctuations is
\begin{equation}
  \partial F_{|(|\delta x\rangle,|\delta p \rangle) = 0} = \left(\begin{array}{cc} S & 0 \\ 0 & S \end{array} \right), \qquad S = \left(\begin{array}{cc} 0 & -I_N \\ \id_{3N} & 0 \end{array} \right)
\end{equation}
with the $\mathfrak M_{N \times N}(\mathbb C)$-valued ``inertia matrix''
\begin{equation}
  I_N = \left(\begin{array}{ccc} Y_2^2+Y_3^2 & -Y_2Y_1+[Y_1,Y_2] & -Y_3Y_1 + [Y_1,Y_3] \\ -Y_1Y_2 + [Y_2,Y_1] & Y_1^2+Y_3^2 & -Y_3Y_2 + [Y_2,Y_3] \\ -Y_1 Y_3 + [Y_3,Y_1] & -Y_2 Y_3 + [Y_3,Y_2] & Y_1^2+Y_2^2 \end{array} \right)
\end{equation}
Let $|\delta x_i \rangle = \delta x_i |u_i \rangle$ be a decomposition of the fluctuation vectors with $\delta x_i$ being the complex amplitude of the quantum fluctuations and $|u_i \rangle$ being the ``polarisation'' of the fluctuations (with $\|u_i(t=0)\|=1$). We set that $\delta x_i$ obeys to the same equation that in the case $N=1$ (with an abelianization of the brane coordinates):
\begin{eqnarray} \label{eqsd2}
  \delta \ddot x_i & = & - \tr |X-x|^2 \delta x_i + \tr \left((X^j-x^j)(X_i-x_i)\right)\delta x_j \nonumber \\
  & & \quad - 2 \overline{\delta x^j} \delta x_j \delta x_i + 2 \overline{\delta x_i} \delta x^j \delta x_j
\end{eqnarray}
It follows that the Jacobian matrix of this equation is the same than for $N=1$ with the inertia matrix:
\begin{equation}
  I = \left(\begin{array}{ccc} \tr(Y_2^2+Y_3^2) & -\tr(Y_1Y_2) & -\tr(Y_1Y_3) \\ -\tr(Y_1Y_2) & \tr(Y_1^2+Y_3^2) & -\tr(Y_2Y_3) \\ -\tr(Y_1 Y_3) & -\tr(Y_2 Y_3) & \tr(Y_1^2+Y_2^2) \end{array} \right)
\end{equation}
The phase trajectory of $\delta x$ is then wrapped around a 6-torus $\mathbb T^6$.\\
Concerning the polarisation, it obeys to
\begin{eqnarray} \label{pola}
  |\ddot u_i \rangle & = & -\mathrm{trm}\,|X-x|^2 |u_i \rangle - 2 \frac{d\ln \delta x_i}{dt} |\dot u_i \rangle \nonumber \\
  & & \quad -\mathrm{trm} \left([X_i,X^j]-(X^j-x^j)(X_i-x_i) \right) \frac{\delta x_j}{\delta x_i} |u_j\rangle \nonumber \\
  & & \quad +\left(\overline{\delta x^j} \delta x_j - 2 \frac{\overline{\delta x_i}}{\delta x_i} \delta x^j \delta x_j \right) |u_i \rangle \nonumber \\
  & & \quad +\left(2 \frac{\overline{\delta x_i}}{\delta x_i} \delta x^j \delta x_j \langle u_i|u^j \rangle - \overline{\delta x^j}\delta x_j \langle u^j|u_i \rangle\right) |u_j \rangle
\end{eqnarray}
where $\mathrm{trm}\, A = A - \tr A$.\\

\begin{figure}
  \begin{center}
    \includegraphics[width=6.4cm]{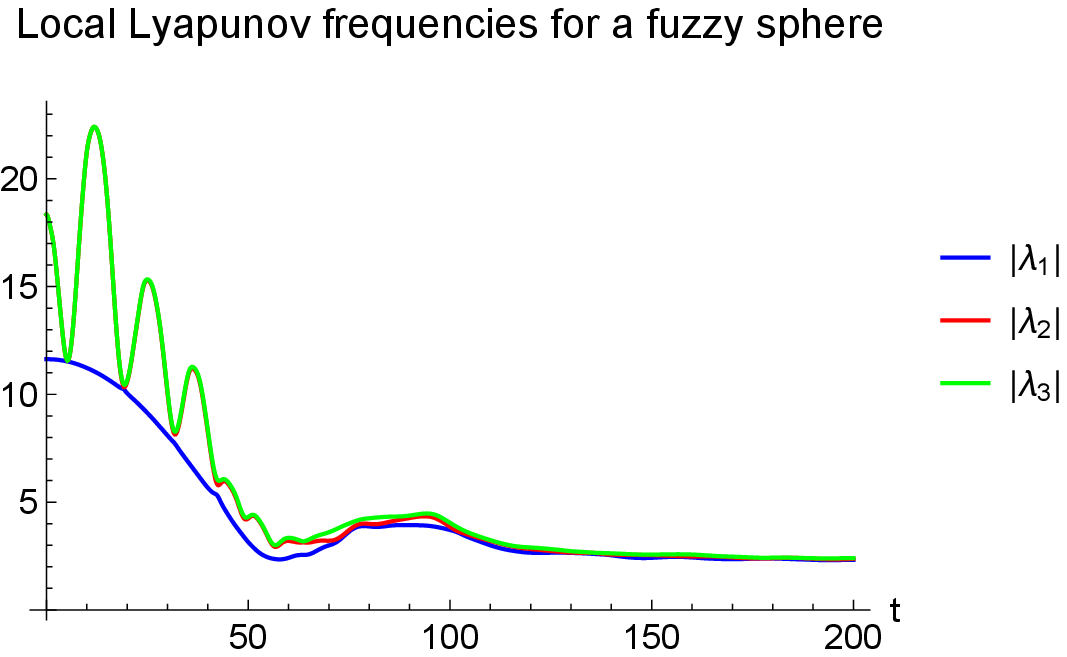} \includegraphics[width=6.4cm]{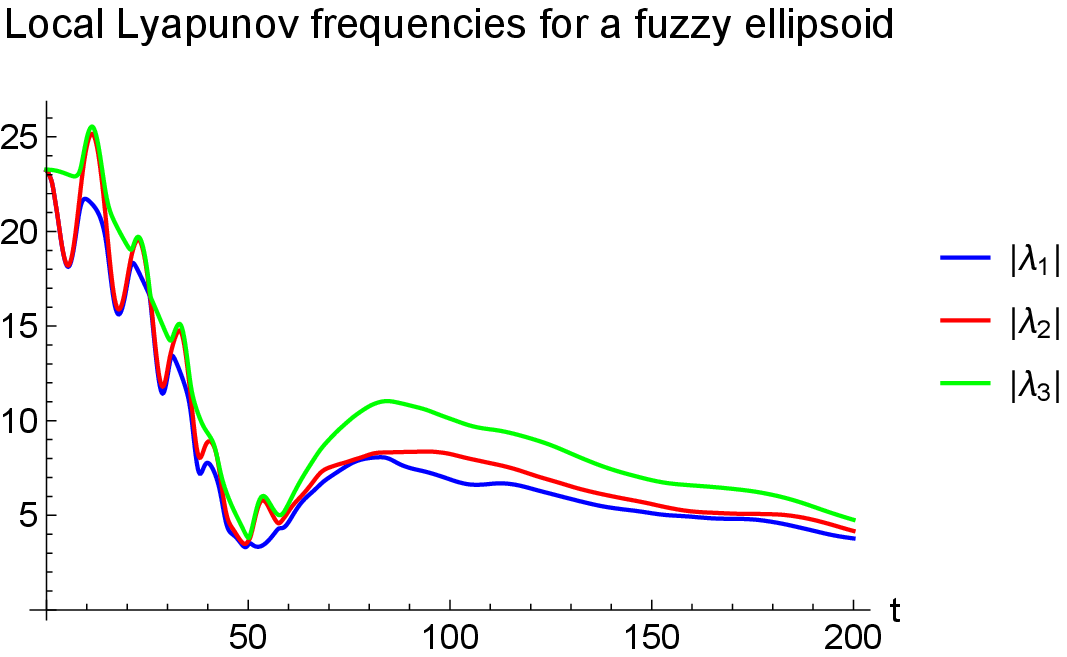}\\
    \includegraphics[width=6.4cm]{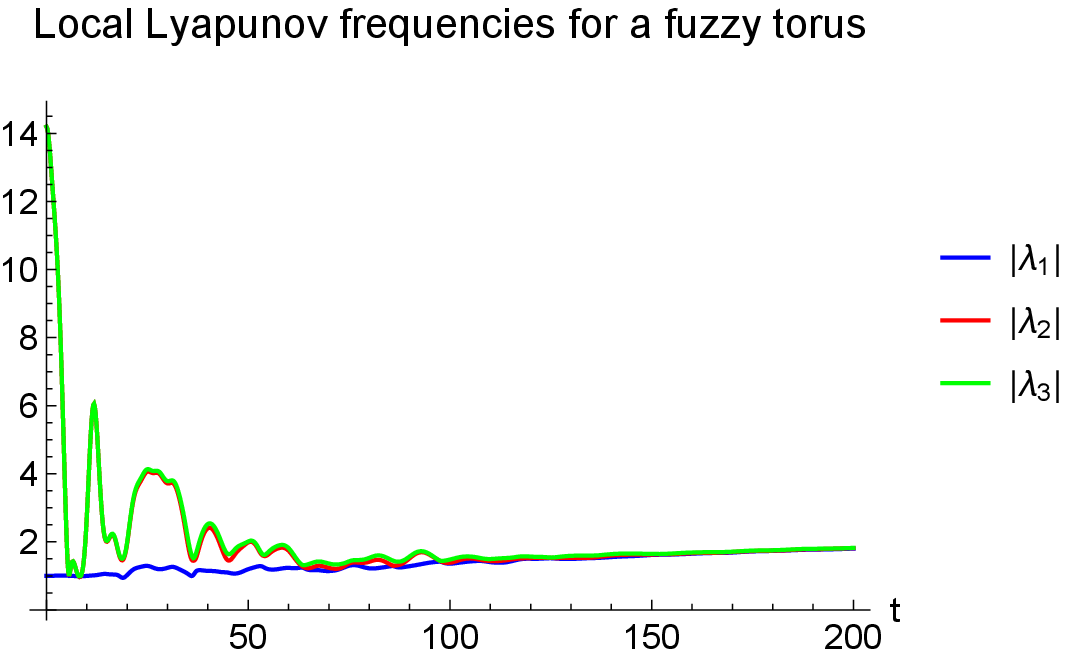} \includegraphics[width=6.4cm]{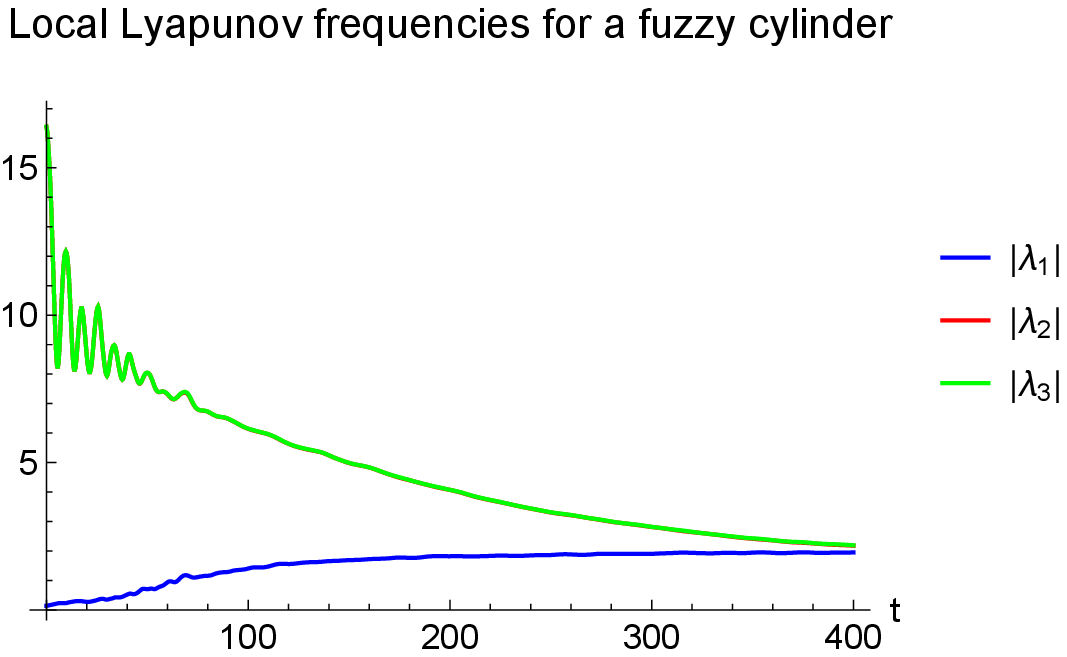}
    \caption{\label{loclyap} Evolution of the local Lyapunov frequencies of the equation \ref{eqsd2} ($\Sp(\sqrt{I})$) for fuzzy sphere, ellipsoid, torus and cylinder. The initial fluctuations are choosen following a gaussian law with $\sigma=0.01$. The number of D0-branes is $N=201$.}
  \end{center}
\end{figure}

In order to illustrate the deformation of the invariant torus with respect to the evolution of $X(t)$ and $x(t)$, we can draw the Lyapunov frequencies of the dynamical system with respect to $t$, see fig. \ref{loclyap}.
The effect of the thermalization can be also viewed with the evolution of the Lyapunov frequencies.

\subsection{Dynamics onto the invariant torus}
Now we want describe the dynamics of the fluctuations onto the invariant torus $\mathbb T^6$. The linearized dynamics is defined by
\begin{eqnarray}
  \delta x(t) & \simeq & \sum_{\alpha=1}^3 (c_\alpha e^{\lambda_\alpha t} \mathring e_\alpha + d_\alpha e^{-\lambda_\alpha t} \mathring e_\alpha^*) \\
  \delta p(t) & \simeq & \sum_{\alpha=1}^3 (c_\alpha e^{\lambda_\alpha t} \hat e_\alpha + d_\alpha e^{-\lambda_\alpha t} \hat e_\alpha^*)
\end{eqnarray}
where $\lambda_\alpha \in \imath \mathbb R$ are the Lyapunov values and $-I \hat e_\alpha = \lambda_\alpha^2 e_\alpha$, $\mathring e_\alpha = \lambda_\alpha \hat e_\alpha$ ($S e_\alpha = \lambda_\alpha e_\alpha$ and $S e_\alpha^* = - \lambda_\alpha e_\alpha^*$, the star denoting the complex conjugation).
\begin{eqnarray}
  c_\alpha & = & \mathring e_\alpha^* \cdot \delta x(0) + \hat e_\alpha^* \cdot \delta p(0) \\
  d_\alpha & = & \mathring e_\alpha \cdot \overline{\delta x(0)} + \hat e_\alpha \cdot \overline{\delta p(0)}
\end{eqnarray}
Let $\{\theta_\alpha\}_{\alpha \in \{1,...,6\}}$ be the local coordinates onto $\mathbb T^6$. Since the previous equations represent a phase trajectory wrapped around $\mathbb T^6$, the immersion equations of $\mathbb T^6$ into $\mathbb C^6$ are
\begin{eqnarray}
  \delta x & \simeq & \sum_{\alpha=1}^3 (c_\alpha e^{\imath \theta_\alpha} \mathring e_\alpha + d_\alpha e^{\imath \theta_{\alpha+3}} \mathring e_\alpha^*) \\
  \delta p & \simeq & \sum_{\alpha=1}^3 (c_\alpha e^{\imath \theta_\alpha} \hat e_\alpha + d_\alpha e^{\imath \theta_{\alpha+3}} \hat e_\alpha^*)
\end{eqnarray}

We rewrite these equations as
\begin{eqnarray}
  \delta x_i & = & \sum_{\alpha=1}^6 \mathring r_{i\alpha} e^{\imath \theta_\alpha} \\
  \delta p_i & = & \sum_{\alpha=1}^6 \hat r_{i\alpha} e^{\imath \theta_\alpha}
\end{eqnarray}
with $\mathring r_{i\alpha} = c_\alpha \mathring e_{\alpha i}$ if $\alpha \leq 3$, $\mathring r_{i\alpha} = d_{\alpha-3} \mathring e_{\alpha-3, i}^*$ if $\alpha > 3$. It follows that
\begin{equation}
  \delta \dot p_i = \imath \sum_{\alpha=1}^6 \hat r_{i\alpha} e^{\imath \theta_\alpha} \dot \theta_\alpha
\end{equation}
Finally we write $\delta \dot p_i = \imath \hat R \dot \theta$ where the matrix $\hat R$ is defined by the elements $\hat r_{i\alpha} e^{\imath \theta_\alpha}$. We have then $\dot \theta = - \imath \hat R^{-1} \delta \dot p$. The dynamics onto $\mathbb T^6$ is then defined by the equation
\begin{equation}
  \dot \theta = \IM \hat R(\theta)^{-1} \hat F(\chi(\theta))
\end{equation}
where $\chi : \mathbb T^6 \to \mathbb C^3$ is defined by $\chi(\theta) = \sum_{\alpha=1}^3 (c_\alpha e^{\imath \theta_\alpha} \mathring e_\alpha + d_\alpha e^{\imath \theta_{\alpha+3}} \mathring e_\alpha^*)$, and $\hat F:\mathbb C^3 \to \mathbb C^3$ is defined by $\hat F_i(\delta x) = - \tr |X-x|^2 \delta x_i + \tr \left((X^j-x^j)(X_i-x_i)\right)\delta x_j - 2 \overline{\delta x^j} \delta x_j \delta x_i + 2 \overline{\delta x_i} \delta x^j \delta x_j$.\\
Finally we have $\dot \theta = F(\theta)$ with
\begin{equation} \label{eqF}
  F = \IM \hat R^{-1} \hat F \circ \chi
\end{equation}

Remark: note that at the linear approximation, $\theta_\alpha(t) \simeq \lambda_\alpha t$ and $\theta_{\alpha+3}(t) \simeq -\lambda_\alpha t$, it follows then $\theta_\alpha(t) \simeq -\theta_{\alpha+3}(t)$.

\section{Dynamics of the fermionic string}
\subsection{Chaotic behaviour}
\begin{figure}
  \begin{center}
    \includegraphics[width=6.4cm]{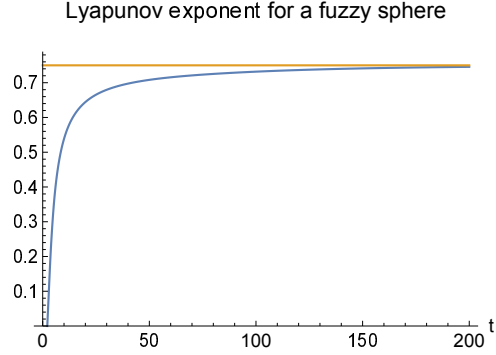}
    \caption{\label{lyap} $\frac{1}{t} \ln \frac{\|X(t)-X_0(t)\|}{\|X(0)-X_0(0)\|}$ (computed with the Sprott algorithm \cite{Gur}) with $X(t)$ and $X_0(t)$ the dynamics of the D2-brane respectively with and without fluctuations in the initial conditions, with $\|X\| = \sqrt{\|X_i\|_2 \|X^i\|_2}$ ($\|\cdot \|_2$ being the usual matrix norm). The system is a fuzzy sphere with the initial fluctuations choosen following a gaussian law with $\sigma=0.01$. The number of D0-branes is $N=201$.
      The curve tends to the dominant Lyapunov exponent $\underline \lambda=0.75$.}
  \end{center}
\end{figure}

We consider now the dynamics of the fermionic string governed by eq. \ref{equSchro} with the Hamiltonian $H^{eff}=\sigma^i \otimes (X_i-x_i)$. The dynamics of the string is indirectly affected by the fluctuations included in the dynamics of $X$ and $x$. Following ref. \cite{Asano, Gur, Hanada}, these dynamics are chaotic. We can verify this fact by studying the dominant asymptotic Lyapunov exponent $\underline \lambda$ of the dynamics of $X$, fig. \ref{lyap}.
$\underline \lambda>0$ confirming its chaotic behaviour. The numerical simulations show that the value of $\underline \lambda$ is highly dependent from the model and from the initial fluctuations. The spin of the fermionic string is then driven by a classical chaotic system.\\

\begin{figure}
  \begin{center}
    \includegraphics[width=6.4cm]{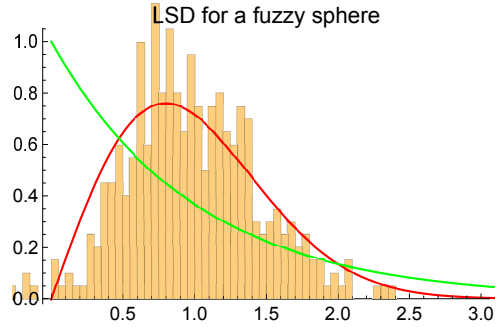}
    \caption{\label{LSD} Level spacing distribution of $H^{eff} = \sigma^i \otimes (X_i-x_i)$ at a time after the thermalization for a fuzzy sphere with the initial fluctuations choosen following a gaussian law with $\sigma=0.01$, compared with the Poisson distribution (green) and with the Wigner-Dyson distribution (red). The number of D0-branes is $N=201$.}
  \end{center}
\end{figure}

The classical chaos in the dynamics of $X(t)$ is established, but we want to know if the quantum Hamiltonian $H^{eff}$ presents quantum chaos. A commonly used criterion of quantum chaos is the level spacing distribution (LSD) of the spectum \cite{Haake}. A regular system presents a LSD as Dirac peaks, a (pseudo)-random system presents a LSD as a Poisson distribution and a chaotic system presents a LSD as a Wigner-Dyson distribution. The LSD for $H^{eff}$ for a fuzzy sphere at a time after thermalization is represented fig. \ref{LSD}.
We see clearly that the LSD of $H^{eff}$ (after thermalization) follows a Wigner-Dyson distribution characteristic of the quantum chaos (we have the same results with the other models). It is interesting to note that the thermalization does not involve a pseudo-random behaviour but well a quantum chaotic behaviour.\\
We have then both classical and quantum chaos: in the classical evolution of $X$ and in the quantum dynamics governed by $H^{eff}$.

\subsection{Horizon of coherence of the reduced density matrix}
\begin{figure}
  \begin{center}
    \includegraphics[width=6.4cm]{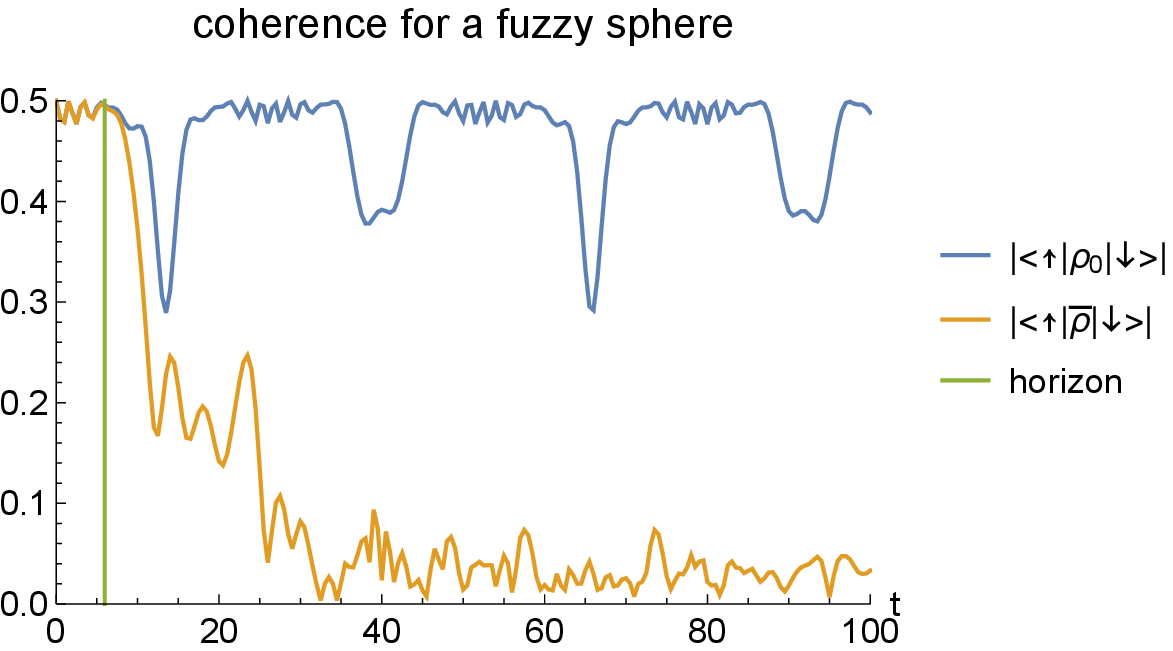} \includegraphics[width=6.4cm]{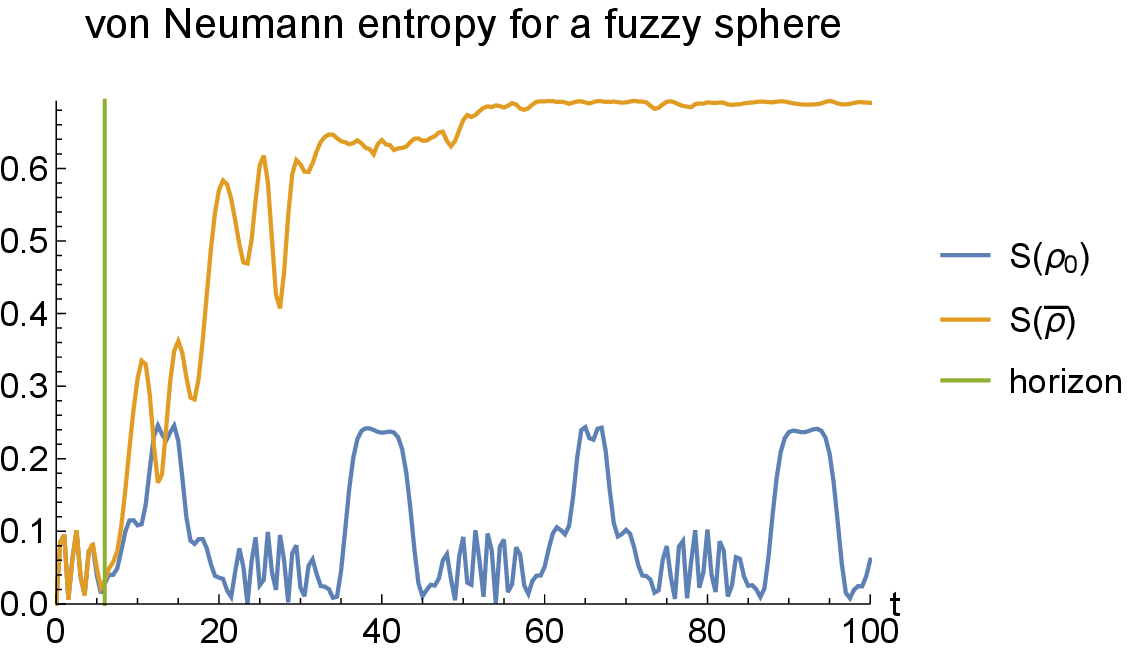}\\
    \includegraphics[width=6.4cm]{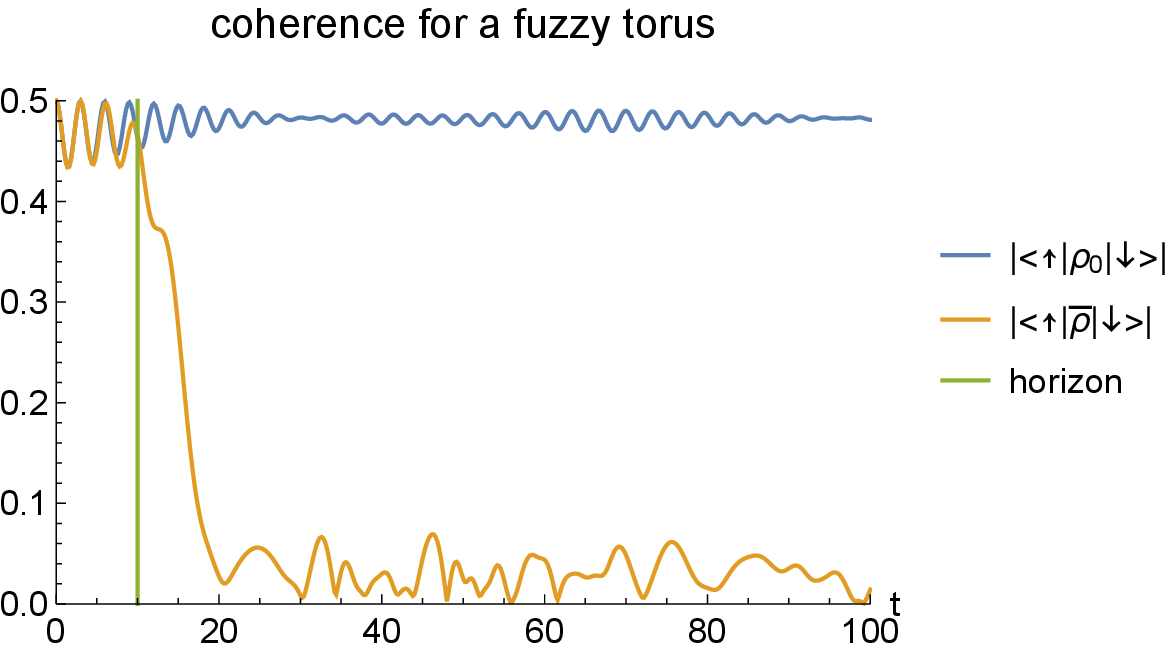} \includegraphics[width=6.4cm]{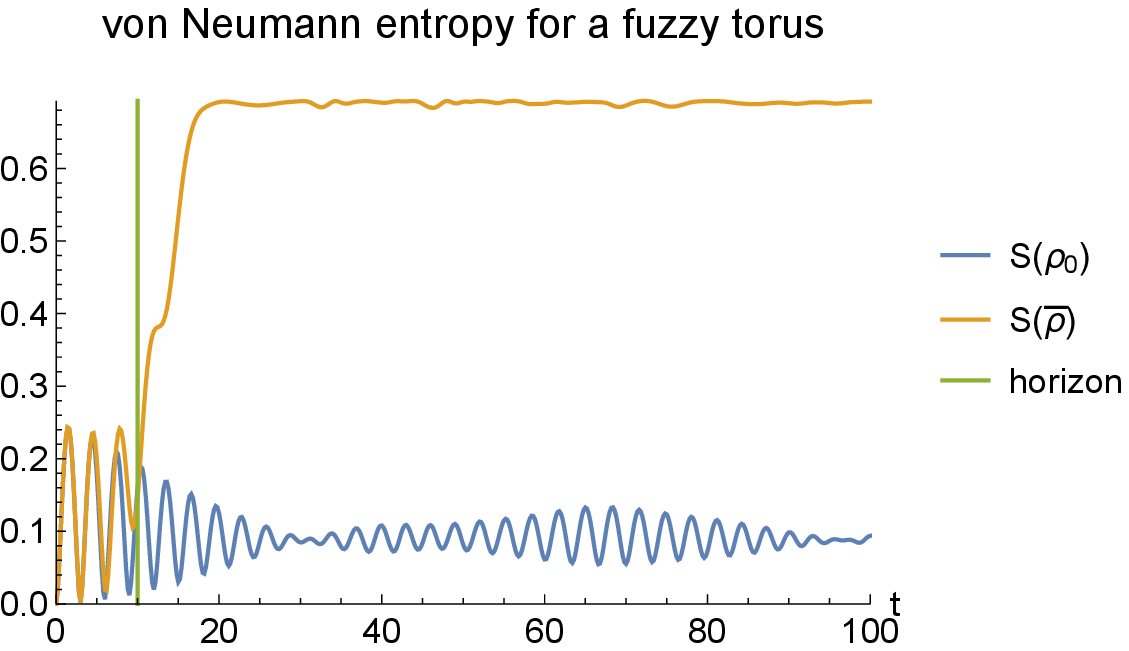}\\
    \includegraphics[width=6.4cm]{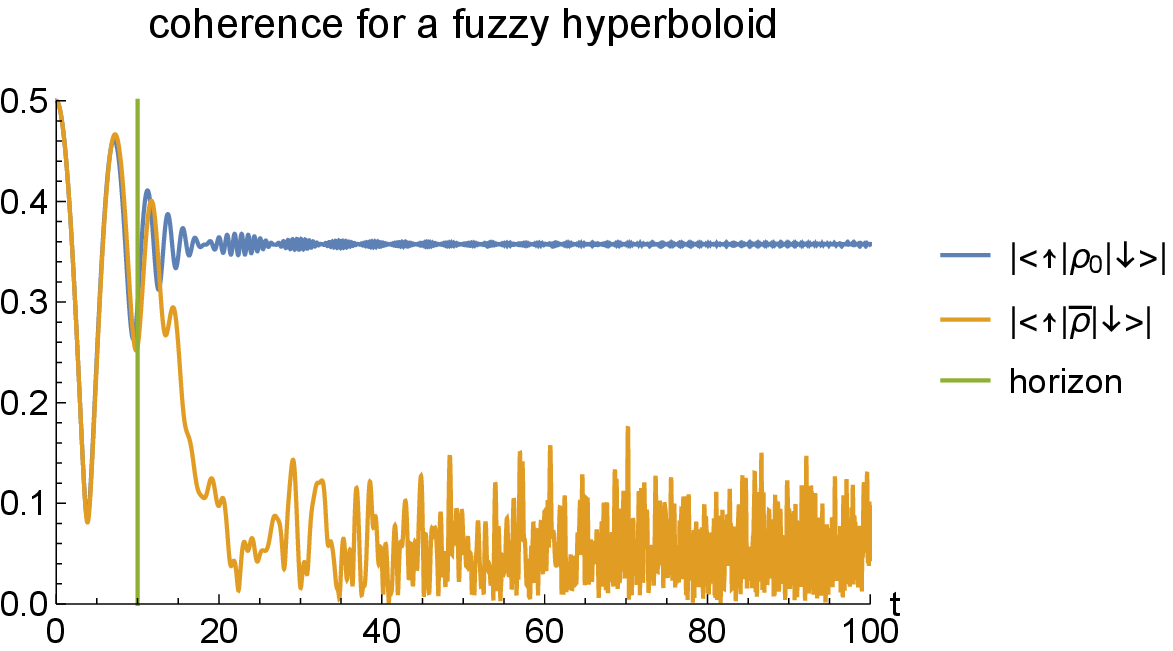} \includegraphics[width=6.4cm]{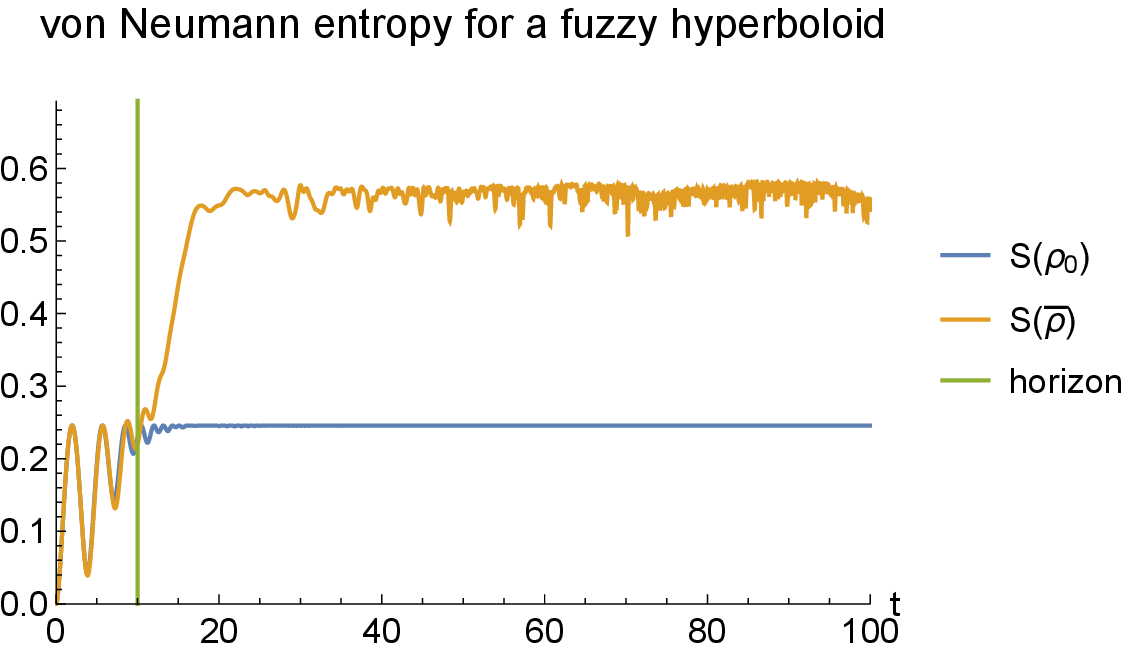}
    \caption{\label{cohe} Coherence and von Neumann entropy of the spin mixed state of the fermionic string, for $X_0$ without fluctuation ($\rho_0$) and for $X$ with fluctuations ($\overline \rho$), with fuzzy sphere, torus and hyperboloid. The initial fluctuations are choosen following a gaussian law with $\sigma=0.01$. The number of D0-branes is $N=11$ (for the fuzzy hyperboloid the cutoff is $N=11$). The vertical line indicates the horizon of coherence. The initial condition for the string is $|\psi(t=0) \rrangle = \frac{1}{\sqrt 2}(|\uparrow \rangle+|\downarrow \rangle) \otimes (1,0,...,0)$.}
  \end{center}
\end{figure}

Let $|\psi \rrangle$ be the state of the fermionic string solution of eq. \ref{equSchro}. We consider the reduced density matrix:
\begin{equation}
  \rho = \tr_{\mathbb C^N} |\psi \rrangle \llangle \psi|
\end{equation}
corresponding to the mixed state of the spin of the string. This spin is subject to the entanglement with the string attachment degree of freedom, to the classical chaotic fluctuations of $X$ and $x$, and to the quantum chaos induced by $H^{eff}$ after the thermalization. In previous works \cite{Viennot2, Aubourg, Aubourg2}, we have shown that a quantum system submitted to a chaotic flow presents an horizon of coherence, an initial duration where the quantum system is preserved from the decoherence phenomenon induced by the chaos. This horizon of coherence is a quantum version of the horizon of predictability of the classical chaotic flows (see discussion in \cite{Viennot2}). In order to study this point for the spin of the fermionic string, we consider the coherence $|\langle \uparrow|\overline \rho|\downarrow \rangle|$ of the averaged density matrix $\overline \rho$ (the average being realized on the initial conditions of the fluctuations) and its von Neumann entropy $S(\overline \rho) = -\tr(\overline \rho \ln \overline \rho)$. The results are compared with the coherence and the von Neumann entropy of the reduced density matrix $\rho_0$ for $X_0$ without fluctuation, in order to distinguish the effects due to the quantum entanglement (present in the two cases) from the effects of the chaos (present only with the fluctuations). Some results are drawn figure \ref{cohe}.
We see the existence of the horizon of coherence before the fall of the coherence and the increase of the entropy, in accordance with the chaotic behaviour. Note that we have the same thing for $\rho$ (without average on the initial fluctuations) but with strongly noisy curves. In contrast with the quantum systems studied in ref. \cite{Viennot2}, we do not see any correlation between the horizon of coherence and the dominant asymptotic Lyapunov exponent. This fact is maybe due to the possibility that the decoherence is dominated by the quantum chaotically behaviour of $H^{eff}$ rather than by the classical chaotically behaviour of $X(t)$, even if the horizon of coherence is shorter than the thermalization duration. In fact, after a lot of numerical simulations with each model of D2-brane and for different values of $\sigma$ (initial dispersion of the fluctuations), we have noted that for the small values of $\sigma$, the horizon of coherence follows a law of the form $t_H = - \alpha \ln \sigma - \beta$ (with $\alpha,\beta >0$ depending on the D2-brane model), and for the large values of $\sigma$, it follows a law of the form $t_H = \frac{\gamma}{\sqrt \sigma} + \delta$ (with $\gamma>0$, $\delta \in \mathbb R$ depending on the D2-brane model). These laws are empirical, a more theoretical analysis needs a better understanding of the quantum chaos or of the intertwining between classical and quantum chaos, which is not the subjet of this paper.

\subsection{Schr\"odinger-Koopman approach and emergent extradimensions}
\begin{figure}
  \begin{center}
    \includegraphics[width=6.4cm]{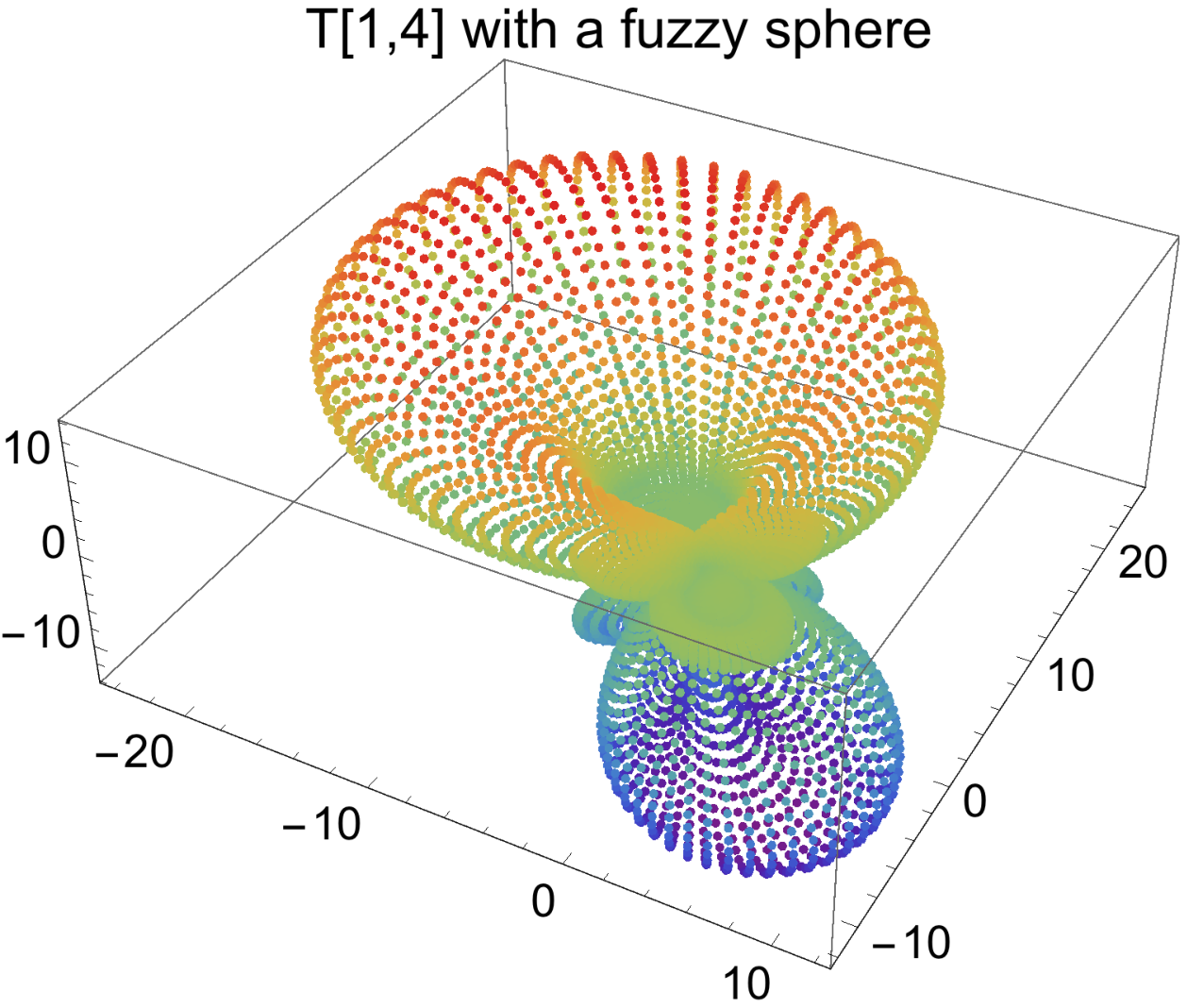} \includegraphics[width=6.4cm]{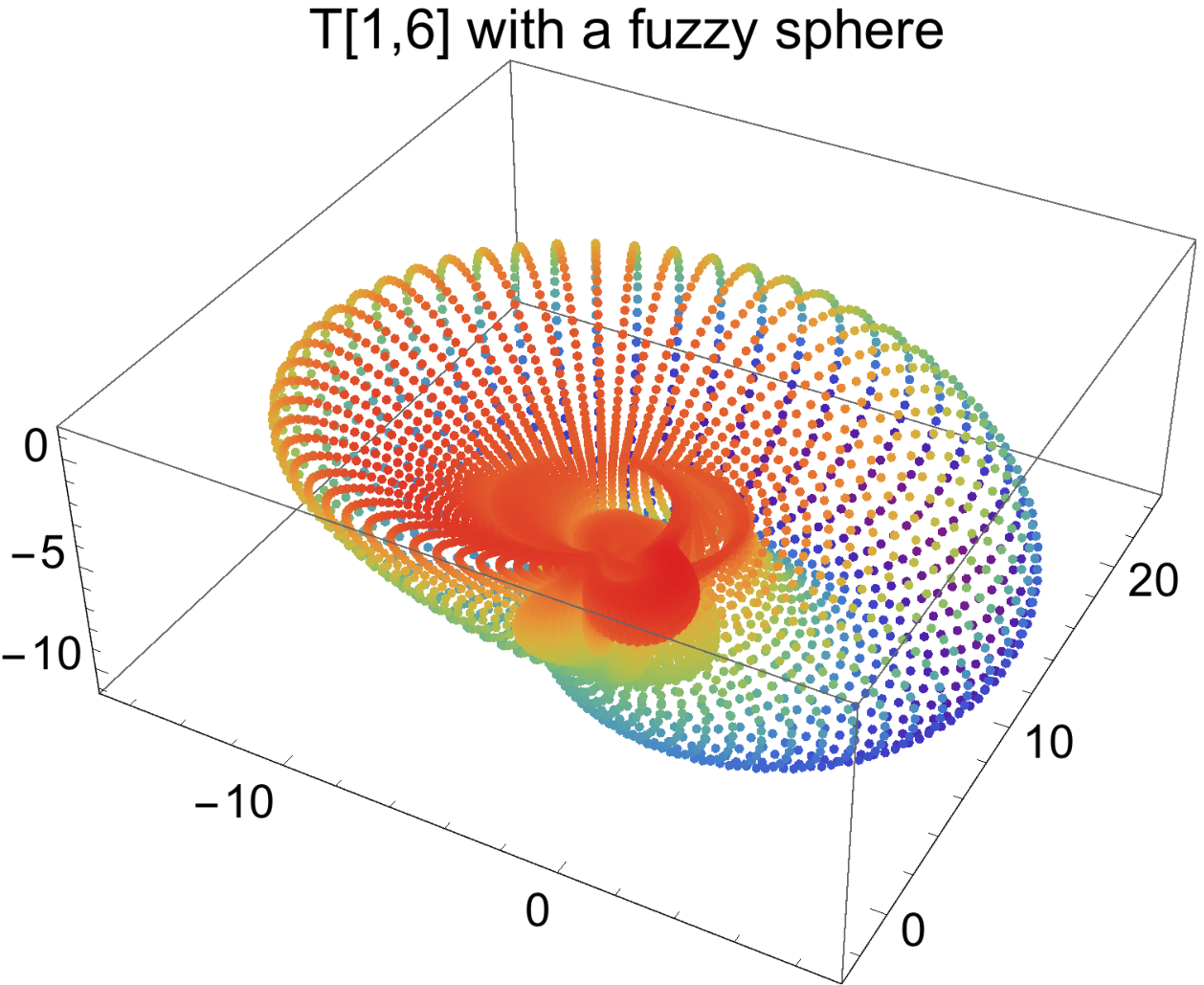}\\
    \includegraphics[width=6.4cm]{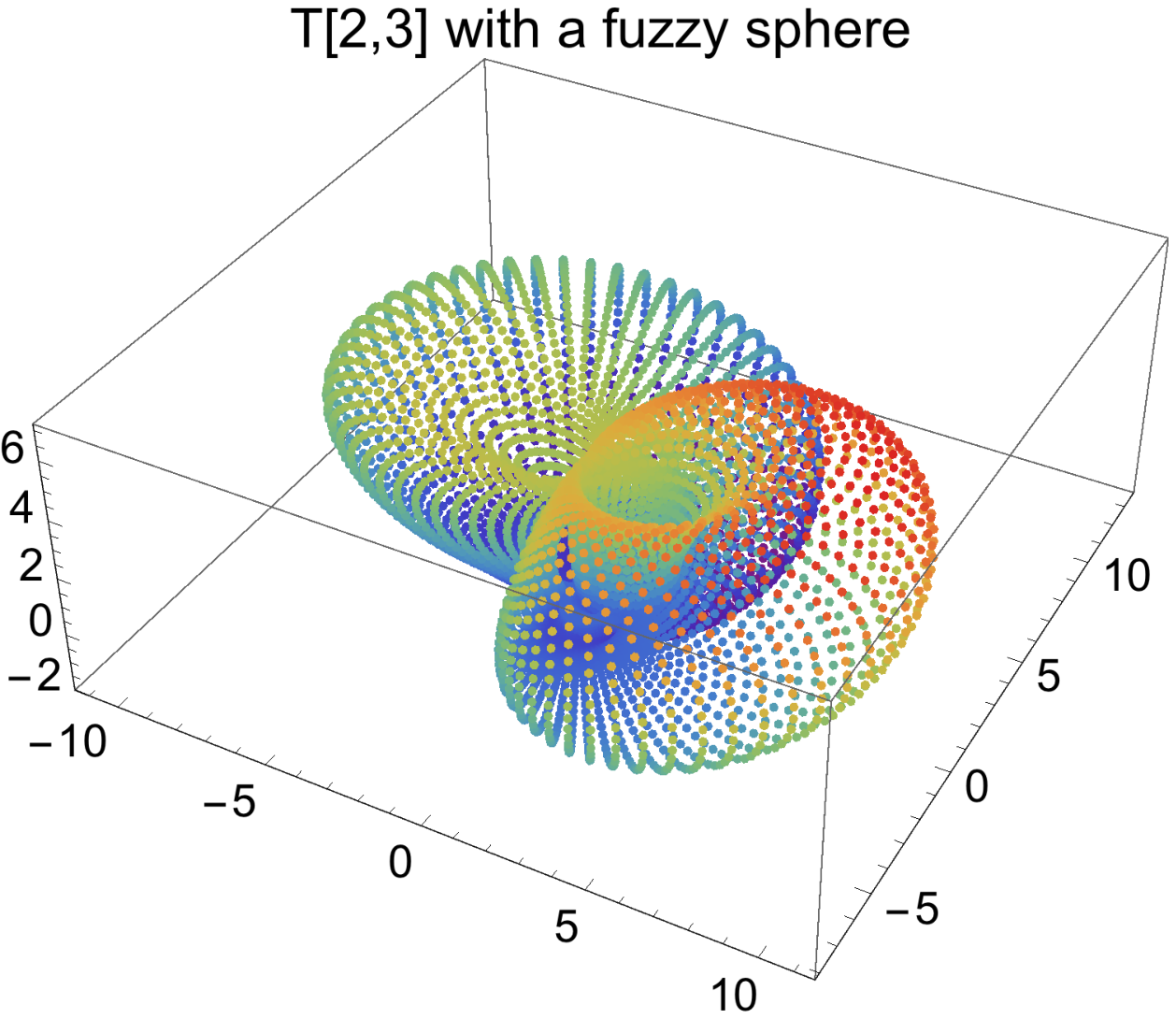} \includegraphics[width=6.4cm]{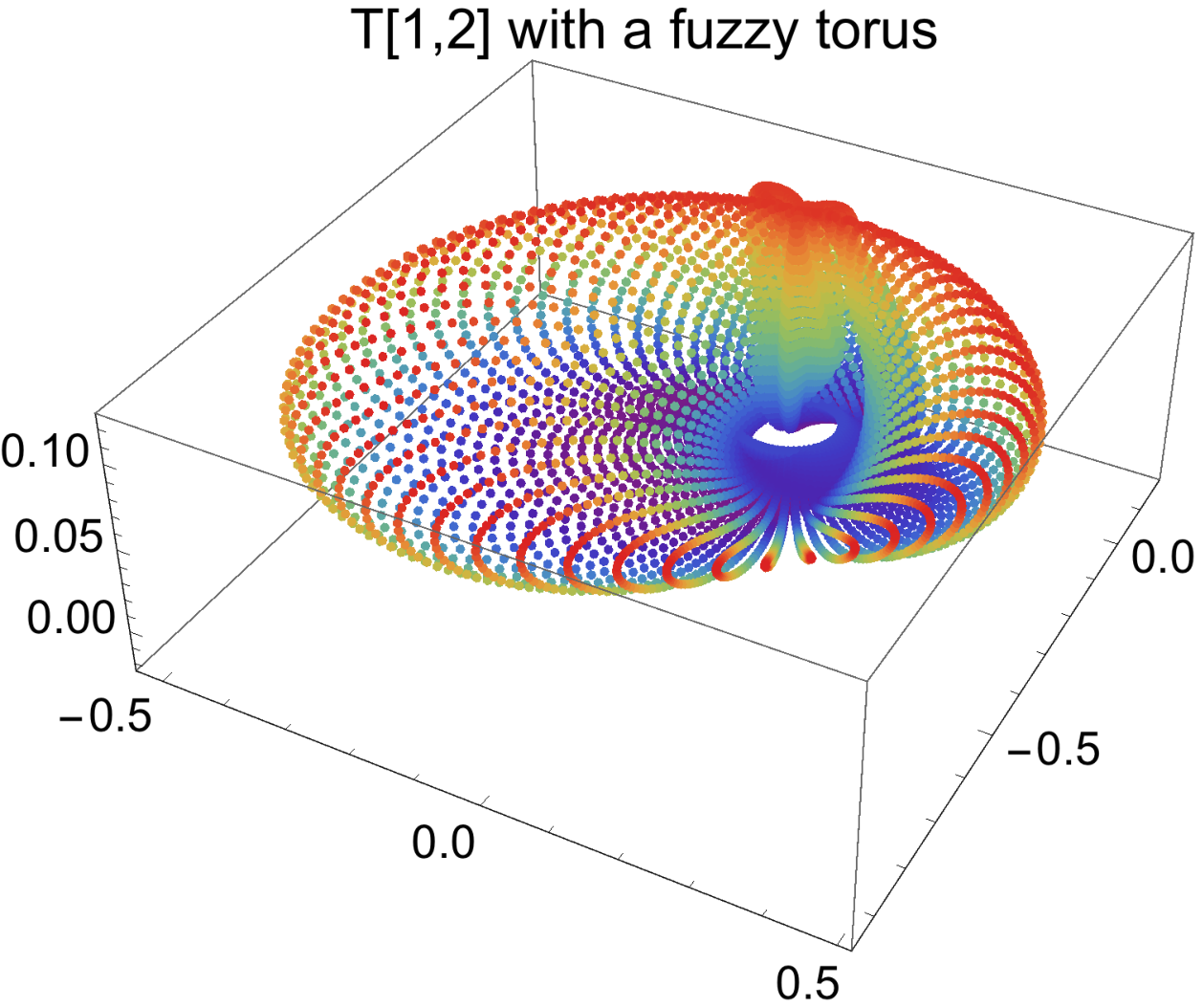}\\
    \includegraphics[width=6.4cm]{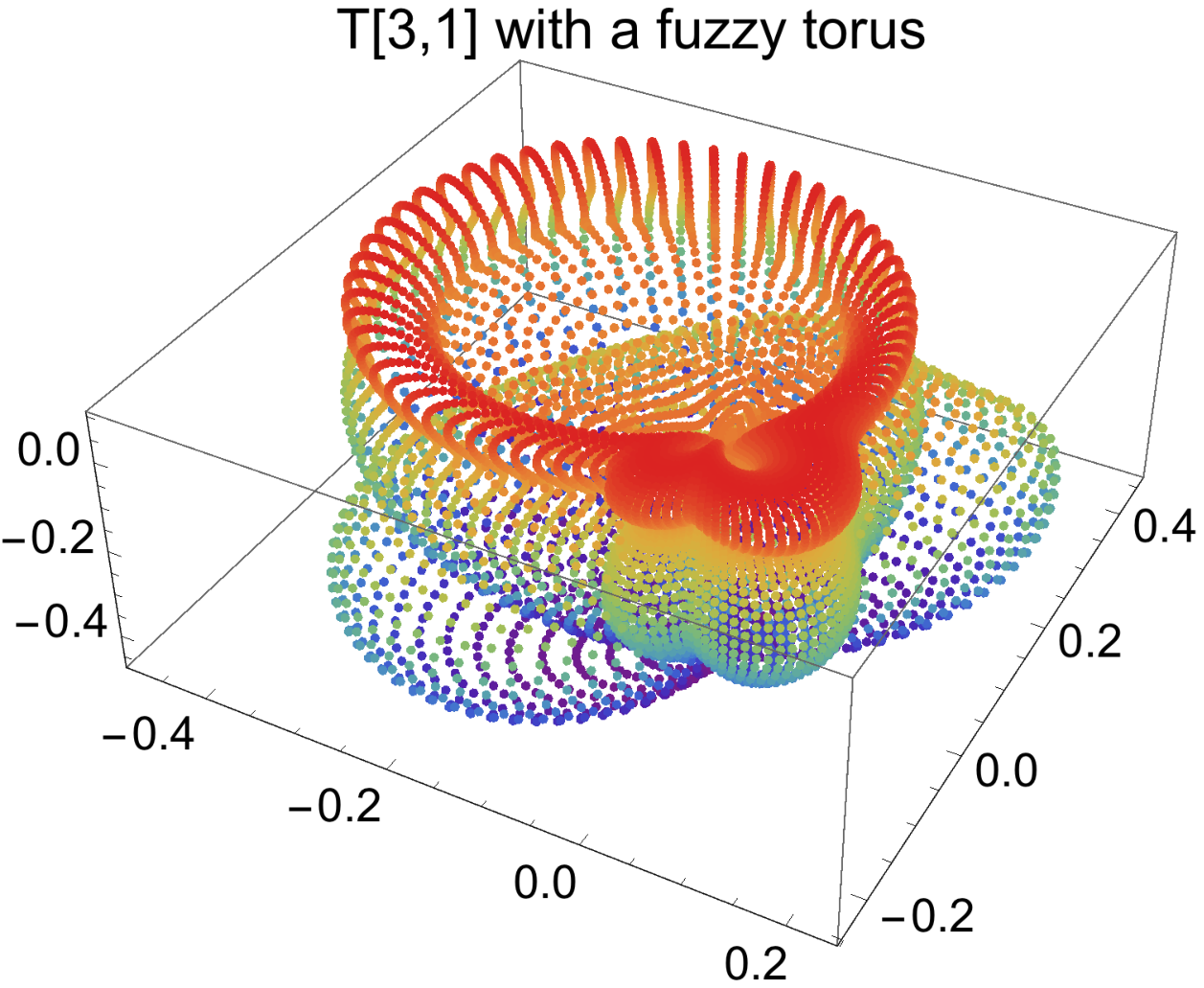} \includegraphics[width=6.4cm]{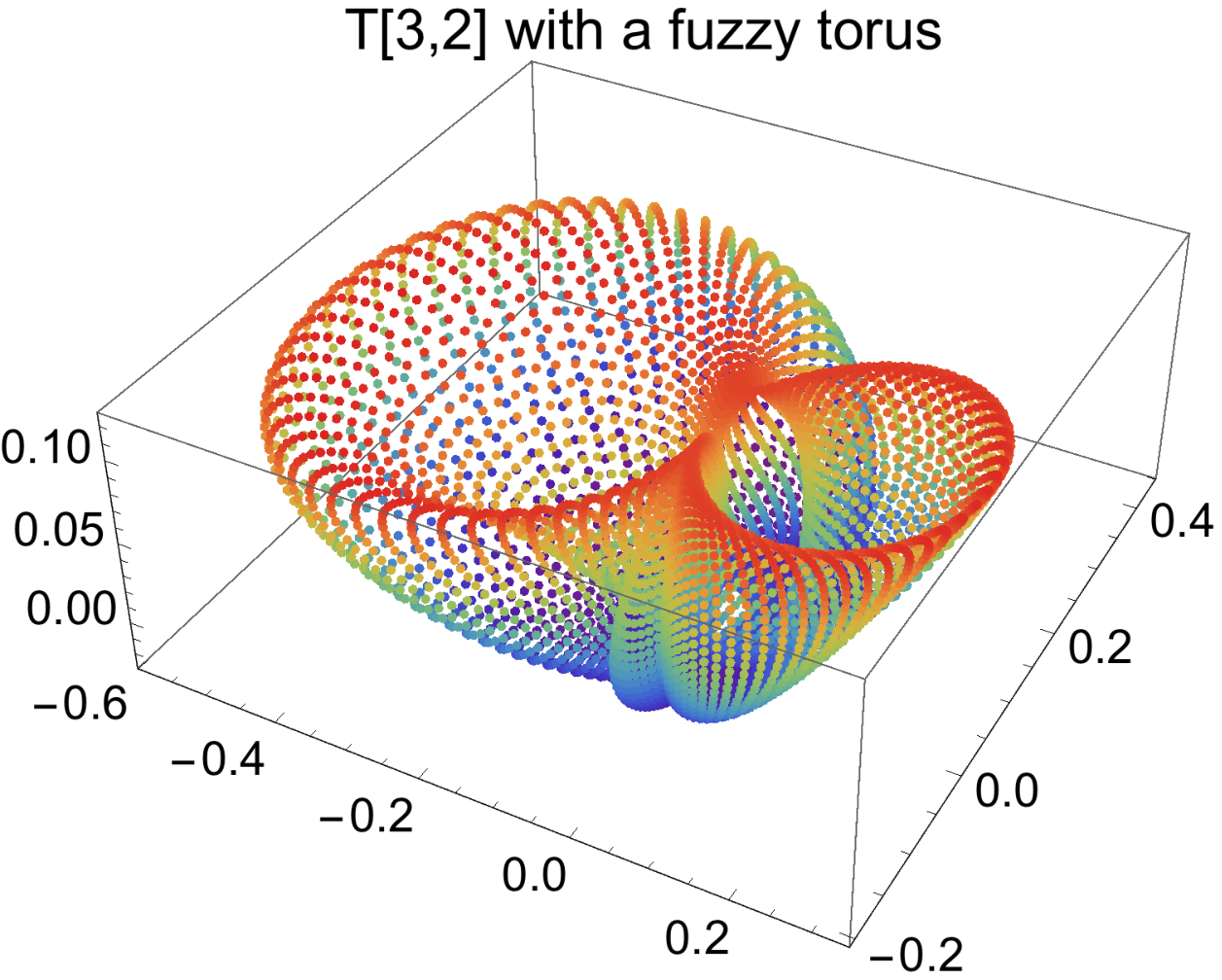}
    \caption{\label{extradim} The compactified dimensions as a non-regular 6-torus of radii $F^\alpha(t,\theta)$ (for $t$ after the thermalization) by section $\theta^\alpha=0$ for four values of $\alpha$. The section 2-torus $T[\alpha,\beta]$ (with $\alpha$ and $\beta$ the indices of the non-fixed angles) is represented by the classical embedding of a torus into $\mathbb R^3$. The initial fluctuations are choosen following a gaussian law with $\sigma=0.01$. The number of D0-branes is $N=11$. The evolution of the geometry is low after the thermalization.}
  \end{center}
\end{figure}

The natural mathematical structure to study a quantum system driven by a chaotic flow is the Schr\"odinger-Koopman approach \cite{Viennot3}. It consists to enlarge the Hilbert space of the quantum system by $L^2(\Gamma,d\mu(q))$ the space of square functions on the stable phase space $\Gamma$ of the chaotic flow ($\mu$ is a probability measure on $\Gamma$ preserved by the flow). In this enlarged Hilbert space, the dynamics is governed by a Koopman Hamiltonian $H_K = -\imath F^\mu(q) \frac{\partial}{\partial q^\mu} \otimes \id + H(q)$ where $H$ is the usual quantum Hamiltonian of the driven system and $F:\Gamma \to \Gamma$ is the generator of the flow. The wave function solution of the Schr\"odinger-Koopman equation $\Psi(t,q)$ and the wave function solution of the usual Schr\"odinger equation $\psi_{q_0}(t)$ are related by $\psi_{q_0}(t) = \Psi(t,q(t))$ with $t\mapsto q(t) \in \Gamma$ the phase trajectory of the chaotic system with $q(0)=q_0$. The interest of the approach is that the mixed state associated with the average onto the initial conditions, $\overline \rho(t) = \int_\Gamma |\psi_{q}(t) \rangle \langle \psi_q(t)| d\mu(q)$, is obtained as a reduced density matrix $\overline \rho(t)= \tr_{L^2(\Gamma,d\mu)} |\Psi(t)\rrangle \llangle \Psi(t)|$. The decoherence induced by the chaotic flow can be then interpreted as an entanglement between the quantum system and the classical chaotic flow.\\
In this section, we want to show how it is possible to adapt the Schr\"odinger-Koopman approach to the fermionic string driven by the D2-brane. Because of section \ref{invtorus} we know that the stable phase space for the chaotic flow can be chosen as being a 6-torus $\mathbb T^6$. The polarization part of the fluctuations is directly driven by the flow onto $\mathbb T^6$ via eq. \ref{pola}. Let $X_i(t,\theta)$ be the D2-brane coordinates for fixed values of $\theta$ ($X(t) = X(t,\theta(t))$ with $\dot \theta = F(\theta)$, $F: \mathbb T^6 \to \mathbb T^6$ being defined eq. \ref{eqF}). Since at the linear limit we have $\theta_\alpha(t) \simeq - \theta_{\alpha+3}(t)$ we have $X(t,..\theta_\alpha..\theta_{\alpha+3}..) \simeq X(t,..-\theta_{\alpha+3}..-\theta_\alpha..)$. Due to this exchange symmetry, we can write that the fermionic state is of the form:
\begin{equation}
  \psi(t) = \psi_{+}(t,\theta_1(t)..\theta_{4}(t)..) + \psi_{-}(t,\theta_{4}(t)..\theta_1(t)..) 
\end{equation}
It follows that
\begin{eqnarray}
  \imath \frac{d}{dt} |\psi(t) \rrangle & = & \imath \partial_t |\psi_+\rrangle + \imath F^1 \partial_1 |\psi_+ \rrangle  + \imath F^4 \partial_4 |\psi_+ \rrangle \nonumber \\
  & & + \imath \partial_t |\psi_-\rrangle + \imath F^4 \partial_1 |\psi_- \rrangle  + \imath F^1 \partial_4 |\psi_- \rrangle
\end{eqnarray}
The dynamics can be rewritten as the following Schr\"odinger-Koopman equation:
\begin{eqnarray}
 && \imath \frac{\partial}{\partial t} \left(\begin{array}{c} |\psi_+\rrrangle \\ |\psi_-\rrrangle \end{array}\right)  =  \left(- \imath \left(\begin{array}{cc} F^1 & 0 \\ 0 & F^4 \end{array} \right) \partial_1 - \imath \left(\begin{array}{cc} 0 & F^1 \\ F^4 & 0 \end{array} \right) \partial_4 \right. \nonumber \\
  & & \left. + \left(\begin{array}{cc} 0 & \sigma^i \otimes (X_i(t,\theta)-x_i(t,\theta)) \\ \sigma^i \otimes (X_i(t,\theta)-x_i(t,\theta)) \end{array} \right) \right) \nonumber \\
  & & \qquad \qquad \qquad \qquad \times \left(\begin{array}{c} |\psi_+\rrrangle \\ |\psi_-\rrrangle \end{array}\right)
\end{eqnarray}
with $|\psi_\pm(t,\theta_{1/4},\theta_2(t),\theta_3(t),\theta_{4/1},\theta_5(t),\theta_6(t)) \rrrangle \in \mathbb C^2 \otimes \mathbb C^N \otimes L^2(\mathbb T^2_{1,4},d\mu)$ (where $d\mu$ is the Haar measure on $\mathbb T^2_{1,4}$ the torus generated by $\theta_1$ and $\theta_4$).\\
$|\psi(t) \rrangle = |\psi_+(t,\theta_1(t),\theta_2(t),\theta_3(t),\theta_4(t),\theta_5(t),\theta_6(t)) \rangle + |\psi_-(t,\theta_4(t),\theta_2(t),\theta_3(t),\theta_1(t),\theta_5(t),\theta_6(t)) \rangle$ is solution of the fermionic string Schr\"odinger equation.\\
Let the deformed Dirac matrices be
\begin{eqnarray}
  \gamma^i & = & \left(\begin{array}{cc} 0 & \sigma_i \\ \sigma_i & 0 \end{array} \right) \\
  \kappa^4 & = & \left(\begin{array}{cc} F^1 & 0 \\ 0 & F^4 \end{array} \right) \simeq F^1 \left(\begin{array}{cc} 1 & 0 \\ 0 & -1 \end{array} \right) \\
  \kappa^7 & = & \left(\begin{array}{cc} 0 & F^1 \\ F^4 & 0 \end{array} \right) \simeq F^1 \left(\begin{array}{cc} 0 & 1 \\ -1 & 0 \end{array} \right)
\end{eqnarray}
where ``$\simeq$'' denotes the linear approximation where $F^4=-F^1$ ($\dot \theta^4 \simeq - \dot \theta^1$). Let $\gamma^4 = \left(\begin{array}{cc} 1 & 0 \\ 0 & -1 \end{array} \right)$ and $\gamma^7 = \left(\begin{array}{cc} 0 & 1 \\ -1 & 0 \end{array} \right)$, it is easy to verify that $\{\gamma^A,\gamma^B\} = \pm 2 \delta^{AB}$ ($\forall A,B=1,...,4,7$). We can then consider that $\gamma^i$, $\kappa^4/F^1$ and $\kappa^7/F^1$ constitute a deformed Clifford algebra (the deformation being associated with the nonlinear part of the dynamics onto the torus).\\
We can now reiterate this argumentation with the couple $(\theta_2,\theta_5)$ by writting that $\psi_\pm(t) = \psi_{\pm +}(t,..\theta_2(t)..\theta_5(t)..) + \psi_{\pm -}(t,..\theta_5(t)..\theta_2(t)..)$ and with $(\theta_3,\theta_6)$ by writting that $\psi_{\pm \pm}(t) = \psi_{\pm \pm +}(t,..\theta_3(t)..\theta_6(t)) + \psi_{\pm \pm -}(t,..\theta_6(t)..\theta_3(t))$.\\
Finally the dynamics is described by the following Schr\"odinger-Koopman equation:
\begin{equation}
  \imath \frac{\partial}{\partial t} |\Psi(t) \rrrangle = \left(-\imath \kappa^{\alpha+3}(t,\theta) \frac{\partial}{\partial \theta^\alpha} + \gamma^i \otimes (X_i(t,\theta)-x_i(t,\theta)) \right) |\Psi(t) \rrrangle
\end{equation}
where $\theta \in \mathbb T^6$, $|\Psi \rrrangle = \mathbb C^{16} \otimes \mathbb C^N \otimes L^2(\mathbb T^6,d\mu)$ (where $d\mu$ is the Haar measure on $\mathbb T^6$). The spinor is defined by \small
\begin{equation}
  \Psi = \left(\begin{array}{c} \psi_{+++} \\ \psi_{++-} \\ \psi_{+-+} \\ \psi_{+--} \\ \psi_{-++} \\ \psi_{-+-} \\ \psi_{--+} \\ \psi_{---} \end{array} \right)
\end{equation}
with $\psi_{\pm\pm\pm} \in \mathbb C^2 \otimes \mathbb C^N \otimes L^2(\mathbb T^6,d\mu)$. The deformed Dirac matrix (generating a deformed Clifford algebra) are defined by
\begin{equation}
  \gamma^i = \left(\begin{array}{ccc} & & \sigma^i \\ & \rotatebox{90}{$\ddots$} & \\ \sigma^i & & \end{array} \right)
\end{equation}
\begin{equation}
  \kappa^4 = \left(\begin{array}{cccccccc} & & & & & & F^1 & \\  & & & & & & & F^4 \\  & & & & F^1 & & & \\  & & & & & F^4 & & \\  & & F^1 & & & & & \\  & & & F^4 & & & & \\ F^1 & & & & & & & \\  & F^4 & & & & & & \end{array} \right)
\end{equation}
\begin{equation}
  \kappa^7 = \left(\begin{array}{cccccccc}  & & & & & & & F^1 \\  & & & & & & F^4 & \\  & & & & & F^1 & & \\  & & & & F^4 & & & \\  & & & F^1 & & & & \\  & & F^4 & & & & & \\  & F^1 & & & & & & \\ F^4 & & & & & & & \end{array} \right)
\end{equation}
\begin{equation}
  \kappa^5 = \left(\begin{array}{cccccccc} & & & & F^2  & & & \\  & & & & & F^2  & & \\  & & & & & & F^5 & \\  & & & & & & & F^5 \\ F^2 & & & & & & & \\  & F^2 & & & & & & \\  & & F^5 & & & & & \\  & & & F^5 & & & & \end{array} \right)
\end{equation}
\begin{equation}
  \kappa^8 = \left(\begin{array}{cccccccc}  & & & & & & & F^2 \\  & & & & & & F^2 & \\ & & & & & F^5 & & \\  & & & & F^5  & & & \\  & & & F^2  & & & & \\  & & F^2 & & & & & \\  & F^5 & & & & & & \\ F^5 & & & & & & & \end{array} \right)
\end{equation}
\begin{equation}
  \kappa^6 = \left(\begin{array}{cccccccc} F^3 & & & & & & & \\  & F^3 & & & & & & \\  & & F^3 & & & & & \\  & & & F^3 & & & &
    \\  & & & & F^6 & & & \\  & & & & & F^6 & & \\  & & & & & & F^6 & \\  & & & & & & & F^6 \end{array} \right)
\end{equation}
\begin{equation}
  \kappa^9 = \left(\begin{array}{cccccccc}  & & & & & & & F^3 \\  & & & & & & F^3 & \\  & & & & & F^3 & & \\  & & & & F^3 & & & \\  & & & F^6 & & & & \\  & & F^6 & & & & & \\  & F^6 & & & & & & \\ F^6  & & & & & & & \end{array} \right)
  \end{equation}
\normalsize
We recover the solution of the usual Schr\"odinger equation by $|\psi(t) \rrangle = \sum_{i=1}^8 \langle \theta(t)|\Psi_i(t) \rrrangle$ where $\theta(t)$ is the trajectory on $\mathbb T^6$.\\
We can rewrite the Schr\"odinger-Koopman equation as:
\begin{equation} \label{eqSKD}
  \imath |\dot \Psi \rrrangle = (\gamma^{\alpha+3} \otimes Y_\alpha + \gamma^i \otimes(X_i-x_i)) |\Psi \rrrangle + Z |\Psi \rrrangle
\end{equation}
where $Y_\alpha = - \imath \Delta F_{\alpha \mod 3} \partial_\alpha$ with $\Delta F_\alpha = \frac{F_\alpha - F_{\alpha+3}}{2}$, and where $Z = - \imath \Sigma^{\alpha \mod 3} \beta^\alpha \partial_\alpha$ modelizes the interaction with the nonlinear fluctuation field, $\Sigma^\alpha = \frac{F^\alpha+F^{\alpha+3}}{2}$ and $\beta^1 = \sigma^1 \otimes \sigma^1 \otimes \sigma^0$, $\beta^2 = \sigma^1 \otimes \sigma^0 \otimes \sigma^0$, $\beta^3 = \sigma^0 \otimes \sigma^0 \otimes \sigma^0$ and $\beta^4=\beta^5=\beta^6=\sigma^1 \otimes \sigma^1 \otimes \sigma^1$.\\
Equation \ref{eqSKD} (without $Z$) is the equation for a fermionic string in a 9-dimensional space $\mathbb R^3 \times \mathbb T^6$ with a D2-brane wrapped in the six compactified dimensions, $\{X_i\}_{i=1,2,3}$ being its noncommutative coordinates in the non-compactified dimensions and $\{Y_\alpha\}_{\alpha = 1,...,6}$ being its noncommutative coordinates in the compactified dimensions (see \cite{Taylor, Brace, Konechny}). The compactified dimension radii are $\frac{F^\alpha(t,\theta)}{2\pi}$ (the geometry is a dynamical variable). It is interesting to note that with the Schr\"odinger-Koopman representation, the six supplementary compactified dimensions can be considered as emerging from the quantum fluctuations in 3-dimensional space without adding any assumption or consideration. Since the space-time geometry is revealed only by test particle (probe fermionic string), it is not possible to distinguish the 4-dimensional space-time with quantum fluctuation in the Schr\"odinger-Koopman picture from a 10-dimensional space-time in the Schr\"odinger picture having the same operator algebra \{$X^i, Y^\alpha \}_{i=1,2,3; \alpha=1,...,6}$.\\
Since the dimension radii are known we can compute the geometry of the extradimensions by the numerical simulations. Some examples are drawn fig. \ref{extradim}.

\section{Conclusion}
The classical and quantum chaotic behaviours of the fluctuations in the D2-brane dynamics involves decoherence in the spin mixed state of a linked fermionic string. This system can be viewed as a model of qubit interacting with a quantum (micro) black hole (the noncommutative D2-brane representing its quantum horizon). Except the duration of the thermalization, the results seems weakly dependent on the number $N$ of D0-branes in the stack, but for a better discussion concerning (large) black holes, it needs to study the thermodynamic limit $N \to +\infty$. But this needs different numerical approaches that the ones used in this paper. The chaotic behaviour induces the existence of an horizon of coherence in the evolution of the spin mixed state (its values seems depend only on the D2-brane model and on the initial dispersion of the fluctuations). The application of the Schr\"odinger-Koopman approach to treat the effects of the fluctuations onto the fermionic string state, makes appear six compactified extradimensions in the modelization. A 4-dimensional space-time with quantum fluctuations in the Schr\"odinger picture is then equivalent to a 10-dimensional space-time in the Schr\"odinger-Koopman picture with fluctuations corresponding to the nonlinear part of the evolution ($Z$ operator in equation \ref{eqSKD}). It would be interesting to generalize the present study to the other matrix theories (BMN and IKKT models \cite{Sochichiu,Zarembo}).

\ack
The authors acknowledge support from I-SITE Bourgogne-Franche-Comt\'e under grants from the I-QUINS project, and support from OSU THETA under grants from the SRO projects. Simulations have been executed on computers of the Utinam Institute supported by the R\'egion de Bourgogne-Franche-Comt\'e and the Institut des Sciences de l'Univers (INSU).

\appendix
\section{About the numerical simulations} \label{AppA}
For the CCR algebra, the numerical simulations need to restrict the description to finite dimensional Hilbert space. Let $N \in \mathbb N^*$ be the cutoff in description and $P_N = \sum_{n=0}^N |n\rangle \langle n|$ be the projector onto the subspace used in the simulations. Let $a_N = P_N a P_N$, $a_N^+ = P_N a^+ P_N$ and $1_N = P_N$ be the numerical representations of the generators of the CCR algebra. The problem is that these operators do not satisfy the definition of the CCR, since
\begin{equation}
  a_N a_N^+ - a_N^+ a_N = 1_{N-1} - N |N \rangle \langle N|
\end{equation}
We see that the error is not small, it is equal to the cutoff value. If it is limited to the last state at this stage, during the propagation of the brane equation, it quickly contaminates all states. The solution is a redefinition of the commutator in the numerical representation:
\begin{equation}
 \forall A,B \in \Env(\mathfrak{ccr}), \quad [A_N,B_N]_N = R_N^{CCR} \circledast (A_N B_N - B_N A_N)
\end{equation}
where $A_N = P_N A P_N$, $\circledast$ is the term to term multiplication : $(A \circledast B)_{ij} = A_{ij} B_{ij}$, and the renormalisation matrix being
\begin{equation}
  (R_N^{CCR})_{ij} = \left\{ \begin{array}{cl} - \frac{1}{N} & \text{if } i=j=N \\ 1 & \text{everywhere} \end{array} \right.
\end{equation}
We have then
\begin{equation}
  [a_N,a_N^+]_N = 1_N
\end{equation}

The same thing occurs for the $\mathfrak{su}(1,1)$ algebra: let $P_N = \sum_{m=0}^N |k,m\rangle \langle k,m|$
\begin{equation}
  K_{N}^+ K_{N}^- - K_{N}^- K_{N}^+ = -2 K_{N-1}^3 + N(N+2k-1)|k,N\rangle \langle k,N|
\end{equation}
and $K_{N}^3 K_{N}^\pm - K_{N}^\pm K_{N}^3 = \pm K_{N}^\pm$. We set then
\begin{eqnarray}
  & & \forall A,B \in \Env(\mathfrak{su}(1,1)) \nonumber \\
          & & \qquad [A_N,B_N]_N = R_N^{SU(1,1)} \circledast (A_N B_N - B_N A_N)
\end{eqnarray}
with the renormalisation matrix
\begin{equation}
  (R_N^{SU(1,1)})_{ij} = \left\{ \begin{array}{cl} - \frac{2(N+k)}{N(N+2k-1)} & \text{if } i=j=N \\ 1 & \text{everywhere} \end{array} \right.
\end{equation}
We have then
\begin{equation}
  [K_N^+,K_N^-]_N = -2 K_N^3 \qquad [K_N^3,K_N^\pm]_N = \pm K_N^\pm
\end{equation}


\section*{References}
\begin{thebibliography}{0}
\bibitem{Viennot} Viennot D and Moro O 2017 {\it Class. Quant. Gravity} {\bf 34}, 055005.
\bibitem{Asplund} Asplund C T, Berenstein D and Trancanelli D 2011 {\it Phys. Rev. Lett.} {\bf 107}, 17602.
\bibitem{Berenstein} Berenstein D and Dzienkowski E 2012 {\it Phys. Rev. D} {\bf 86}, 086001.
\bibitem{Aoki} Aoki S, Hanada M and Iizuka N 2015 {\it JHEP} 2015, 029.
\bibitem{BFSS} Banks T, Fischler W, Shenker S H and Susskind L 1997 {\it Phys. Rev. D} {\bf 55}, 5112.
\bibitem{Asano} Asano Y, kawai D and Yoshida K 2015 {\it JHEP} 2015, 191.
\bibitem{Gur} Gur-Ari G, Hanada M and Shenker S H 2016 {\it JHEP} 2016, 091.
\bibitem{Hanada} Hanada M, Shimada H and Tezuka M 2017 {\it preprint} arXiv:1702.06935.
\bibitem{Viennot2} Viennot D and Aubourg L 2013 {\it Phys. Rev. E} {\bf 87}, 062903.
\bibitem{Aubourg} Aubourg L and Viennot D 2015 {\it Quant. Inf. Process.} {\bf 14}, 1117.
\bibitem{Aubourg2} Aubourg L and Viennot D 2016 {\it J. Phys. B} {\bf 49}, 115501.
\bibitem{Viennot3} Viennot D and Aubourg L 2018 {\it preprint} arXiv:1802.08186
\bibitem{Sapin} Sapin O, Jauslin H R and Weigert S 2007 {\it J. Stat. Phys.} {\bf 127}, 699.
\bibitem{Jauslin} Jauslin H R and Sugny D 2010 in {\it Mathematical horizons for quantum physics''} (Singapore: World Scientific).
\bibitem{Gay} Gay-Balmaz F and Tronci C 2018 {\it preprint} arXiv:1802.04787.
\bibitem{Viennot4} Viennot D 2009 {\it J. Phys. A} {\bf 42} 395302.
\bibitem{Sochichiu} Sochichiu C 2006 {\it Lect. Notes Phys.} {\bf 698}, 189.
\bibitem{Zarembo} Zarembo K L and Makeenko Y M 1998 {\it Uspekhi Fizicheskikh Nauk} {\bf 168}, 3.
\bibitem{Badyn} Hudoba de Badyn M, Karczmarek J L, Sabella-Garnier P and Huai-Che Yeh K 2015 {\it JGEP} 2015, 89.
\bibitem{Sykora} Sykora A 2016 {\it preprint} arXiv:1610.015041
\bibitem{Goldstein} Goldstein H, Poole C, and Safko J 2000 {\it Classical Mechanics} (New York: Addison Wesley).
\bibitem{Haake} Haake F 1991 {\it Quantum signature of chaos} (Berlin: Springer-Verlag).
\bibitem{Taylor} Taylor W 1998 {\it preprint} arXiv:hep-th/9801182.
\bibitem{Brace} Brace D, Morariu B and Zumino B 1999 {\it Nucl. Phys. B} {\bf 545}, 192.
\bibitem{Konechny} Konechny A and Schwarz A 2002 {\it Phys. Rept.} {\bf 360}, 353.
\end{thebibliography}
\end{document}